\PassOptionsToPackage{unicode}{hyperref}
\PassOptionsToPackage{hyphens}{url}
\PassOptionsToPackage{dvipsnames,svgnames,x11names}{xcolor}
\documentclass[
  12pt]{article}

\usepackage{amsmath,amssymb}
\usepackage{iftex}
\usepackage[T1]{fontenc}
\usepackage[utf8]{inputenc}
\usepackage{textcomp} 
\usepackage{lmodern}
\usepackage{xcolor}
\makeatletter
\ifx\paragraph\undefined\else
  \let\oldparagraph\paragraph
  \renewcommand{\paragraph}{
    \@ifstar
      \xxxParagraphStar
      \xxxParagraphNoStar
  }
  \newcommand{\xxxParagraphStar}[1]{\oldparagraph*{#1}\mbox{}}
  \newcommand{\xxxParagraphNoStar}[1]{\oldparagraph{#1}\mbox{}}
\fi
\ifx\subparagraph\undefined\else
  \let\oldsubparagraph\subparagraph
  \renewcommand{\subparagraph}{
    \@ifstar
      \xxxSubParagraphStar
      \xxxSubParagraphNoStar
  }
  \newcommand{\xxxSubParagraphStar}[1]{\oldsubparagraph*{#1}\mbox{}}
  \newcommand{\xxxSubParagraphNoStar}[1]{\oldsubparagraph{#1}\mbox{}}
\fi
\makeatother

\usepackage{longtable,booktabs,array}
\usepackage{calc} 
\usepackage{etoolbox}
\makeatletter
\patchcmd\longtable{\par}{\if@noskipsec\mbox{}\fi\par}{}{}
\makeatother
\IfFileExists{footnotehyper.sty}{\usepackage{footnotehyper}}{\usepackage{footnote}}
\makesavenoteenv{longtable}
\usepackage{graphicx}
\graphicspath{{figures/}}
\makeatletter
\def\maxwidth{\ifdim\Gin@nat@width>\linewidth\linewidth\else\Gin@nat@width\fi}
\def\maxheight{\ifdim\Gin@nat@height>\textheight\textheight\else\Gin@nat@height\fi}
\makeatother
\setkeys{Gin}{width=\maxwidth,height=\maxheight,keepaspectratio}
\makeatletter
\def\fps@figure{htbp}
\makeatother

\makeatletter
\@ifpackageloaded{caption}{}{\usepackage{caption}}
\AtBeginDocument{%
\ifdefined\contentsname
  \renewcommand*\contentsname{Table of contents}
\else
  \newcommand\contentsname{Table of contents}
\fi
\ifdefined\listfigurename
  \renewcommand*\listfigurename{List of Figures}
\else
  \newcommand\listfigurename{List of Figures}
\fi
\ifdefined\listtablename
  \renewcommand*\listtablename{List of Tables}
\else
  \newcommand\listtablename{List of Tables}
\fi
\ifdefined\figurename
  \renewcommand*\figurename{Figure}
\else
  \newcommand\figurename{Figure}
\fi
\ifdefined\tablename
  \renewcommand*\tablename{Table}
\else
  \newcommand\tablename{Table}
\fi
}
\@ifpackageloaded{float}{}{\usepackage{float}}
\floatstyle{ruled}
\@ifundefined{c@chapter}{\newfloat{codelisting}{h}{lop}}{\newfloat{codelisting}{h}{lop}[chapter]}
\floatname{codelisting}{Listing}

\makeatother
\makeatletter
\@ifpackageloaded{caption}{}{\usepackage{caption}}
\@ifpackageloaded{subcaption}{}{\usepackage{subcaption}}
\makeatother

\usepackage[]{natbib}
\usepackage{bookmark}

\IfFileExists{xurl.sty}{\usepackage{xurl}}{} 
\hypersetup{
  pdftitle={Robust Bayesian Inference of Causal Effects via Randomization Distributions},
  pdfauthor={Easton Huch, Fred Feinberg, Walter Dempsey},
  pdfkeywords={Bernstein--von Mises theorem, Fisherian randomization test, model checking, nonparametric method},
  colorlinks=true,
  linkcolor={blue},
  filecolor={Maroon},
  citecolor={Blue},
  urlcolor={Blue},
  pdfcreator={LaTeX via pandoc}}

\newcommand{\anon}{1}

\usepackage{mathtools}
\usepackage{multirow}
\usepackage{enumitem}
\usepackage{multibib}
\newcites{App}{Appendix References}

\usepackage{amsthm}
\theoremstyle{plain}
\newtheorem{theorem}{Theorem}

\newtheorem{lemma}{Lemma}
\newtheorem{corollary}{Corollary}
\newtheorem{example}{Example}
\theoremstyle{definition}
\newtheorem{definition}{Definition}
\newtheorem{assumption}{Assumption}
\theoremstyle{remark}

\renewcommand\vec[1]{\boldsymbol{#1}}
\newcommand\mat[1]{\mathbf{\uppercase{#1}}}

\newcommand{\E}{\mathbb{E}}

\newcommand{\PP}{\text{P}}
\newcommand{\R}{\mathbb{R}}
\newcommand{\N}{\mathbb{N}}

\newcommand{\A}{\mathcal{A}}
\newcommand{\Mtheta}{\mathcal{M}_{\vec{\theta}}}
\newcommand{\indep}{\perp \!\!\! \perp}

\DeclareMathOperator*{\supp}{supp}

\DeclareMathOperator*{\Var}{Var}
\DeclareMathOperator*{\Varhat}{\widehat{Var}}

\DeclareMathOperator*{\plim}{plim}
\DeclareMathOperator*{\HalfNormaldist}{Half-Normal}
\DeclareMathOperator*{\Normaldist}{Normal}
\DeclareMathOperator*{\Gammadist}{Gamma}
\DeclareMathOperator*{\Bernoullidist}{Bernoulli}
\newcommand{\transpose}{{\top}}
\newcommand{\Y}{\mathcal{Y}}

\begin{document}

\def\spacingset#1{\renewcommand{\baselinestretch}%
{#1}\small\normalsize} \spacingset{1}


\if1\anon
{
  \title{\bf Robust Bayesian Inference of Causal Effects via Randomization Distributions}
  \author{
    Easton Huch\thanks{
    This work was supported by the National Institute on Drug Abuse (NIDA) under Grant number P50DA054039 and the National Institute of General Medical Sciences (NIGMS) under Grant number R01GM152549.
    Dr. Huch gratefully acknowledges research funding from the Johns Hopkins Carey Business School that supported him during the late stages of this project.}\hspace{.2cm}\\
    Department of Statistics, University of Michigan\\
    and \\
    Fred Feinberg \\
    Department of Marketing, University of Michigan \\
    and \\
    Walter Dempsey \\
    Department of Biostatistics, University of Michigan
    }
  \maketitle
} \fi

\if0\anon
{
  \bigskip
  \bigskip
  \bigskip
  \begin{center}
    {\LARGE\bf Robust Bayesian Inference of Causal Effects via Randomization Distributions}
\end{center}
  \medskip
} \fi

\bigskip
\begin{abstract}
We present a general framework for Bayesian inference of causal effects that delivers provably robust inferences founded on design-based randomization of treatments.
The framework involves fixing the observed potential outcomes and forming a likelihood based on the randomization distribution of a statistic.
The method requires specification of a treatment effect model;
in many cases, however, it does not require specification of marginal outcome distributions, resulting in weaker assumptions compared to Bayesian superpopulation-based methods.
We show that the framework is compatible with posterior model checking in the form of posterior-averaged randomization tests.
We prove several theoretical properties for the method, including a Bernstein--von Mises theorem and large-sample properties of posterior expectations.
In particular, we show that the posterior mean is asymptotically equivalent to Hodges--Lehmann estimators, which provides a bridge to many classical estimators in causal inference, including inverse-probability-weighted estimators and H\'ajek estimators.
We evaluate the theory and utility of the framework in simulation and a case study involving a nutrition experiment.
In the latter, our framework uncovers strong evidence of effect heterogeneity despite a lack of evidence for moderation effects.
The basic framework allows numerous extensions, including the use of covariates, sensitivity analysis, estimation of assignment mechanisms, and generalization to nonbinary treatments.
\end{abstract}

\noindent%
\vfill

\newpage
\spacingset{1.8} 

\section{Introduction}
Randomization-based causal inference methods offer the promise of valid statistical inference based principally on the physical act of randomization \citep{ding_paradox_2017}.
Randomization-based methods encompass both Neymanian inference \citep{splawa-neyman_application_1990} and Fisherian randomization tests \citep[FRTs;][]{fisher_design_1935}.
Both methods fix potential outcomes at their realized values and use random treatment assignments as the basis for statistical inference, thereby avoiding the need for superpopulation sampling assumptions.
An additional benefit of randomization-based methods is that they position the assignment mechanism as the conceptual focal point of the analysis, facilitating discussion of covariate balance and the risk of hidden confounding---two central issues in applied causal analysis.

The Neymanian approach fixes the full collection of potential outcomes at their realized values and tests the \emph{weak} null hypothesis of no effect on average: $\Bar{y}_0 = \Bar{y}_1$, where $\Bar{y}_j = \sum_{i=1}^n y_{ji} /n$ for $j=0,1$ with $y_{ji}$ denoting the potential outcome for unit $i \in \{1, \ldots, n\}$ under treatment $j$.
FRTs, on the other hand, fix the observed potential outcomes and test the \emph{sharp} null hypothesis of no causal effect for any unit: $y_{0i} = y_{1i}$ for all $i$.
Inference then proceeds by comparing an observed statistic to its randomization distribution under the sharp null.
In both Neymanian inference and FRTs, the stochasticity derives entirely from random treatment assignments; the potential outcomes are fixed.

In contrast, most Bayesian causal inference methods rely on correct specification of outcome models \citep[ch. 8]{rubin_bayesian_1978, Imbens_Rubin_2015}.
Given unconfounded treatment assignment and certain conditions on the prior distribution, the assignment mechanism drops out of the likelihood---a phenomenon that has generated considerable debate in the literature \citep{robins_toward_1997}.
The \emph{ignorability} of the assignment mechanism in these cases has important implications for the robustness of Bayesian causal inference methods.
In particular, Bayesian methods tend to be more sensitive to correct specification of outcome models than their frequentist counterparts because the propensity score does not (in general) balance subject characteristics between treatment and control groups---its primary purpose in most frequentist methods \citep{li_balancing_2018}.

Although Bayesian statisticians largely agree that the assignment mechanism is an important component of a causal analysis, a recent review of Bayesian causal inference concluded that ``there is no consensus on how to proceed'' \citep{li_bayesian_2023}.
Existing strategies include (a) treating the propensity score as a covariate in the outcome model \citep{Rubin1985Propensity}, (b) specifying dependent priors \citep{CHIB200267}, and (c) computing frequency-based point estimators with posterior predictive samples \citep{saarela_bayesian_2016}.
However, these strategies are not universally applicable and raise challenging questions regarding trade-offs among competing analytical priorities, such as robustness to model misspecification, valid uncertainty quantification, and philosophical coherence.

Setting this challenge aside, \citet{li_bayesian_2023} argue that the Bayesian approach offers compelling advantages for causal inference.
First, the Bayesian approach can be applied to a wide variety of causal estimands, even those that do not admit nonparametric large-sample inference, such as individual treatment effects.
Second, Bayesian inferences are automatic in the sense that the inferences---including uncertainty quantification---flow directly from the probabilistic assumptions.
Third, Bayesian inferences offer a simple, straightforward solution for incorporating prior information and pooling inferences across multiple data sources.
Fourth, Bayesian methods are highly extensible and modular.

By placing the assignment mechanism at the center of a Bayesian causal analysis, our proposed framework inherits both the robustness of randomization-based methods \emph{and} the above benefits of the Bayesian paradigm.
We name the resulting framework \emph{Bayesian randomization inference} (BRI) to emphasize the combination of these complementary strengths.
The key idea underlying BRI is to condition on the values of the \emph{observed} potential outcomes.
We then form a statistic that involves model-based imputations of counterfactuals, and we use its randomization distribution as a likelihood function.

Recent work in Bayesian causal inference has begun developing related randomization-based procedures in special settings.
The procedures have predominantly focused on bounded outcomes, such as binary \citep{humphreys_mixing_2015, keele_bayesian_2017, ding_modelfree_2019} or ordinal outcomes \citep{chiba_bayesian_2018}.
To our knowledge, the only exception is the approach of \citet{leavitt_randomization-based_2023}, which can be viewed as a special case of our method with a binary treatment, a constant treatment effect model, and the difference-in-means (DIM) statistic.
Our proposed framework is much more general and can accommodate a wide variety of outcome types, treatment effect models, and statistics.
Our contribution is both the framework itself and the theoretical results of Section \ref{sec:theory} showing that (under certain regularity conditions) BRI models often target nonparametric causal estimands \emph{even if} the Bayesian model is misspecified.

A common feature shared by BRI and Leavitt's approach is that neither is \emph{fully} Bayesian.
In the former case, this feature is the result of BRI decoupling the observed potential outcomes from the assignment vector (see Section \ref{sec:setup}).
In the latter, because Leavitt's approach uses a Gaussian ``working model'' with a robust plug-in variance estimate.
BRI offers a Bayesian alternative to Leavitt's plug-in strategy in the form of posterior model checks \citep{gelman_posterior_1996}.
These checks perform an FRT for each posterior sample, similar to the procedures described in \citet{ding_causal_2018} and \citet{DingGuo2023}.

We introduce the basic BRI framework in Section \ref{sec:framework} and discuss special considerations for discrete statistics in Section \ref{sec:discrete-stats}.
Section \ref{sec:theory} provides the frequentist properties of a large class of BRI models; specifically, we develop a Bernstein--von Mises theorem and show asymptotic equivalence of the posterior mean to Hodges--Lehmann estimators.
Section \ref{sec:case-study} illustrates the framework in an analysis of a nutrition experiment.
Section \ref{sec:discussion} concludes with a discussion of the main results, limitations, and potential extensions of this work.

\section{Basic Framework}
\label{sec:framework}

This section introduces the general framework for BRI.

\subsection{Problem Setup and Assumptions}
\label{sec:setup}

Throughout we use lowercase unbolded characters for scalars ($a, \theta$), lowercase bold characters for vectors ($\vec{a}, \vec{\theta}$), and uppercase bold characters for matrices ($\mat{A}, \mat{\Theta}$).
Because all quantities are potentially random in the Bayesian approach, we do not distinguish between random and fixed/known quantities in the notation, but we clarify this distinction as needed.

We denote treatment assignments as $a_i \in \A \subseteq \R$ for $i \in [n] \coloneqq \{1, 2, \dots, n\}$ with $\vec{a} \in \A^n$ denoting the vector of treatment assignments: $\vec{a} \coloneqq [a_1, \ldots, a_n]^{\transpose}$.
Throughout the main paper, we consider binary treatments with $\A = \{0, 1\}$; 
online Appendix \ref{sec:beyond-binary} discusses the generalization to other treatment types.
We assume the existence of real-valued potential outcomes $y_{0i}, y_{1i} \in \Y \subseteq \R$ for all $i \in [n]$.
Due to the \emph{fundamental problem of causal inference}, we observe only a single potential outcome, $y_{ai}$, for each observation \citep{holland_statistics_1986}.

We denote the vectors of control and treated potential outcomes as $\vec{y_0} \coloneqq [y_{01}, \ldots, y_{0n}]^{\transpose} \in \Y^n$ and $\vec{y_1} \coloneqq [y_{11}, \ldots, y_{1n}]^{\transpose} \in \Y^n$, respectively, with the collection of all potential outcomes denoted as $\mat{Y} \coloneqq [\vec{y_0}\quad \vec{y_1}] \in \R^{n \times 2}$.
The proposed framework involves fixing the observed potential outcomes, $\vec{y_a} \coloneqq [y_{a1}, \ldots, y_{an}]^{\transpose} \in \Y^n$, at their realized values. 
Concretely, if $a_i = 0$, then we fix $y_{0i}$ at its realized value; otherwise, we fix $y_{1i}$ at its realized value.
In both cases, the corresponding potential outcome is the $i$th element of $\vec{y_a}$, so fixing $\vec{y_a}$ effectively fixes half of the potential outcomes at their realized values.
We use $\PP(\cdot)$ and $\PP(\cdot | \cdot)$ to denote the marginal and conditional distributions of their arguments, respectively.
We employ the following causal assumptions:

\begin{assumption}\label{assumption:consistency}
    (\emph{Consistency}) The observed outcomes, $\vec{y}$, are equal to the potential outcomes under the observed treatment assignment: $\vec{y} = \vec{y_a}$.
\end{assumption}

\begin{assumption}\label{assumption:unconfoundedness}
    (\emph{Unconfoundedness}) The treatment assignments are randomly assigned independent of the potential outcomes: $\vec{a} \indep \mat{Y}$.
\end{assumption}

\begin{assumption}\label{assumption:known_assignment_mechanism}
    (\emph{Known Assignment Mechanism}) The random assignment mechanism, $\PP(\vec{a})$, is known.
\end{assumption}

We employ Assumption \ref{assumption:consistency} throughout this article.
In online Appendix \ref{app:extensions}, we outline several generalizations of the basic framework that require weaker versions of Assumptions \ref{assumption:unconfoundedness} and \ref{assumption:known_assignment_mechanism}, such as random assignment given covariates (Assumptions \ref{assumption:conditional-unconfoundedness} and \ref{assumption:conditional_known_assignment_mechanism}).
We impose these strong versions of Assumptions \ref{assumption:unconfoundedness} and \ref{assumption:known_assignment_mechanism} to clarify the exposition.

From the perspective of the Bayesian analyst, we decouple $\vec{y_a}$ from the observed assignment vector, $\vec{a}$, so that $\vec{y_a}$ provides information only on its elements (the observed potential outcomes) but not $\vec{a}$.
For example, suppose $n=4$ and we observe $\vec{a} = (0, 1, 1, 0)^{\transpose}$ and $\vec{y} = (1.2, 4.9, 3.4, 3.6)^{\transpose}$; then the analysis would fix $y_{01} = 1.2$, $y_{12} = 4.9$, $y_{13} = 3.4$, and $y_{04} = 3.6$ but treat $\vec{a}$ as random drawn from the known distribution $\PP(\vec{a})$.
We express this mathematically as $\sigma(\vec{y_a}) \subset \sigma(\mat{Y})$, where $\sigma(\cdot)$ denotes the $\sigma$-field generated by its argument, implying that $\vec{a} \indep \vec{y_a}$ and $\PP(\vec{a}) = \PP(\vec{a} | \vec{y_a})$ by Assumptions \ref{assumption:unconfoundedness} and \ref{assumption:known_assignment_mechanism}, respectively.
This decoupling results in an approximate Bayesian analysis due to the reuse of $\vec{a}$ in both fixing $\vec{y_a}$ and observing the statistic, $\vec{s}$.

\subsection{The Treatment Effect Model}
\label{sec:treatment-effect-model}

BRI requires specification of a treatment effect model $\Mtheta$, indexed by a parameter $\vec{\theta} \in \R^p$, that produces imputations of one or both counterfactuals for each $i$, independently.
The model $\Mtheta$ is a set consisting of parametric forms for $\PP(y_{0i} | y_{1i})$, $\PP(y_{1i} | y_{0i})$, or both; we denote these submodels as $\PP_{\vec{\theta}}(y_{0i} | y_{1i})$ and $\PP_{\vec{\theta}}(y_{1i} | y_{0i})$, respectively.
The form of $\Mtheta$ has important implications for the analysis.
To facilitate the discussion, we introduce the following definitions.

\begin{definition}
\label{definition:directional}
A treatment effect model $\Mtheta$ is \emph{unidirectional} if it contains only one submodel;
otherwise, it is \emph{multidirectional}.
\end{definition}

\begin{definition}
\label{definition:deterministic}
A treatment effect model $\Mtheta$ is \emph{deterministic} if each of its submodels assigns probability one to a single outcome for every value in its conditioning set;
otherwise, it is \emph{stochastic}.
\end{definition}

\begin{definition}
\label{definition:bijective}
A deterministic multidirectional treatment effect model $\Mtheta$ is \emph{bijective} if, for all $\vec{a}, \vec{a}' \in \A^n$, it can be expressed in terms of a bijective function $\vec{m}_{\vec{\theta}}(\cdot, \vec{a}, \vec{a}'): \Y^n \to \Y^n$ such that $\vec{m}_{\vec{\theta}}(\vec{y_a}, \vec{a}, \vec{a}') = \vec{y}_{\vec{a}'}$.
\end{definition}

Throughout, we restrict attention to bijective models and stochastic unidirectional models because the BRI framework provides the greatest benefit for these model types; in particular, we can avoid specifying marginal outcome distributions.
An example of a bijective treatment effect model is the constant treatment effect model $\vec{y}_1 = \vec{y}_0 + \vec{1}_n \theta$, where $\theta \in \R$ and $\vec{1}_n$ is an $n$-vector of ones \citep{rosenbaum_observational_2002}.
In the notation of Definition \ref{definition:bijective}, the constant treatment effect model can be expressed as $\vec{m}_{\vec{\theta}}(\vec{y_a}, \vec{a}, \vec{a'}) = \vec{y_a} + (\vec{a'} - \vec{a}) \theta$.
An example of a stochastic unidirectional model is
\begin{equation}
\label{eq:gaussian-effects}
\PP_{\vec{\theta}}(y_{1i} | y_{0i}) = \text{Normal}(\alpha + y_{0i}, \sigma^2)
\end{equation}
with $\PP(y_{0i} | y_{1i})$ unspecified.
Under \eqref{eq:gaussian-effects}, we have $\vec{\theta} = [\alpha, \sigma]^{\transpose}$.
Although we cannot observe $y_{0i}$ and $y_{1i}$ simultaneously, this model has observable implications.
In particular, it implies that $\E (y_{1i}) = \alpha + \E(y_{0i})$ and $\Var(y_{1i}) = \Var(y_{0i}) + \sigma^2$, provided $\E(y_{0i})$ and $\Var(y_{0i})$ exist.
A generalization of \eqref{eq:gaussian-effects} is
\begin{equation}
\label{eq:gaussian-reg}
\PP_{\vec{\theta}}(y_{1i} | y_{0i}) = \text{Normal}(\alpha + \beta y_{0i}, \sigma^2).
\end{equation}
Under \eqref{eq:gaussian-reg}, we have $\E (y_{1i}) = \alpha + \beta \E(y_{0i})$ and $\Var(y_{1i}) = \beta^2 \Var(y_{0i}) + \sigma^2$, which can accommodate data having $\Var(y_{1i}) \leq \Var(y_{0i})$.
All three models avoid the need to specify marginal distributions for $y_{1i}$ and $y_{0i}$, thereby removing some of the distributional assumptions needed for Bayesian causal inference relative to a superpopulation approach.

\subsection{The Statistic}
\label{sec:statistic}

BRI also requires the analyst to specify a statistic, denoted by $\vec{s} = \vec{f}(\vec{y_a}, \vec{a})$ for a known function $\vec{f}: \R^n \times \A^n \to \R^k$.
Because the analysis is performed conditional on $\vec{y_a}$, the statistic summarizes $\vec{a}$, effectively discarding information in $\vec{a}$ that the analyst considers uninformative (or minimally informative) for the estimation of treatment effects.
BRI is similar to limited-information Bayes (LIB) methods in this respect \citep{kwan_asymptotic_1999, kim_limited_2002}.
The statistic must have a known distribution given $\PP(\vec{a})$, $\vec{y_a}$, $\vec{\theta}$, and the model $\Mtheta$.
For concreteness, consider the statistics
\begin{equation}
\label{eq:treatment-mean}    
s_0 \coloneq
\frac{\sum_{i=1}^n (1 - a_i) y_{ai}}{\sum_{i=1}^n 
(1 - a_i)}, \quad
s_1 \coloneq
\frac{\sum_{i=1}^n a_i y_{ai}}{\sum_{i=1}^n 
a_i}.
\end{equation}
Under models \eqref{eq:gaussian-effects} and \eqref{eq:gaussian-reg}, $s_1$ has a known distribution given $\vec{y_a}$ and $\vec{\theta}$ because these models provide (stochastic) imputations of $\vec{y_1}$.
In contrast, the distribution of $s_0$ is unknown because models \eqref{eq:gaussian-effects} and \eqref{eq:gaussian-reg} do not specify $\PP(y_{0i} | y_{1i})$.
In general, unidirectional treatment effect models require statistics that depend on a single potential outcome: the one imputed by the model.
In contrast, bijective treatment effect models are compatible with statistics involving \emph{both} $\vec{y_0}$ and $\vec{y_1}$, the canonical example being the DIM statistic: $s_{\Delta} \coloneqq s_1 - s_0$.

Although any statistic meeting the above criteria is permissible within the BRI framework, the theoretical results in Section \ref{sec:theory} show that the choice of statistic determines the statistical properties of the resulting posterior distribution.
Whenever practical, we recommend setting $\dim(\vec{s}) \eqqcolon k = p \coloneqq \dim(\vec{\theta})$, selecting elements of $\vec{s}$ that are expected to identify each element of $\vec{\theta}$.
For example, based on the moment calculations for model \eqref{eq:gaussian-effects}, we might specify one element of $\vec{s}$ as $s_1$ and the other as
\begin{equation}
\label{eq:s12}
s_{12} \coloneq
\frac{\sum_{i=1}^n a_i (y_{ai} - s_1)^2}{\sum_{i=1}^n 
a_i}
\end{equation}
to identify the parameters $\alpha$ and $\sigma$, determining the mean and variance of $y_{1i}$.
Section \ref{sec:theory} explores the implications behind the choice of statistic.


\subsection{Model Structure}
\label{sec:model-structure}

Our Bayesian inference procedure involves fixing the potential outcomes $\vec{y_a}$ to their observed values.
This setup is analogous to Bayesian regression models in which the covariates, $\mat{X} \in \R^{n \times q}$, are typically not modeled;
rather, we obtain inferences for model parameters fixing the values of the covariates to their observed values.
In effect, this approach places $\mat{X}$ in every conditioning set (often implicitly) so that the posterior density for the regression parameter $\vec{\beta}$ can be written as
\begin{equation}
\label{eq:bayes-reg}
p(\vec{\beta} | \vec{y}, \mat{X}) \propto
p(\vec{\beta} | \mat{X})\, p(\vec{y} | \vec{\beta}, \mat{X}),
\end{equation}
where $\propto$ denotes proportionality and  $p(\cdot | \cdot )$ represents the conditional density (mass) function of its arguments (throughout the paper, we assume that such density functions exist).
This strategy is often justified in regression modeling by the fact that it avoids imposing unnecessary distributional assumptions on the covariates \citep[p. 354]{gelman_bayesian_2014};
\citet[p. 5]{li_bayesian_2023} provides a related argument in a causal setting.

In a similar fashion, the BRI framework fixes the value of $\vec{y_a}$, effectively placing it in the conditioning set of the prior, likelihood, and posterior as in \eqref{eq:bayes-reg}.
The posterior density is
\begin{equation}
\label{eq:bri-model}
    p(\vec{\theta} | \vec{s}, \vec{y_a}) \propto
    p(\vec{\theta} | \vec{y_a})\, p(\vec{s} | \vec{\theta}, \vec{y_a}),
\end{equation}
where $p(\vec{\theta} | \vec{y_a})$ is the prior density for $\vec{\theta}$ and $p(\vec{s} | \vec{\theta}, \vec{y_a})$ is the likelihood function under the model, $\Mtheta$, and known assignment mechanism, $\PP(\vec{a})$.
The latter is the density function for the randomization distribution of $\vec{s}$ given $\vec{y_a}$ and a fixed value of $\vec{\theta}$.
When $\Mtheta$ is bijective, conditioning on $\vec{\theta}$ provides imputations of the counterfactuals.
In this case, $\vec{a}$ is the sole source of randomness in $p(\vec{s} | \vec{\theta}, \vec{y_a})$, and this probability mass function (PMF) is precisely the PMF that would be used to conduct an FRT for a prespecified value of $\vec{\theta}$ under $\Mtheta$.
For stochastic unidirectional models, an analogous statement holds averaging over random imputations of the counterfactuals.
This connection facilitates model checking in the form of posterior-averaged FRTs; see online Appendix \ref{app:model-checking} for a detailed discussion.

The distribution $\PP(\vec{s} | \vec{\theta}, \vec{y_a})$ is defined as $\PP(\vec{s} | \vec{\theta}, \vec{y_a})
\coloneqq
\Pr\left\{\vec{f}(\vec{y_{\widetilde{a}}}, \widetilde{\vec{a}}) \leq \vec{s} | \vec{\theta}, \vec{y_a}\right\}$, where $\vec{f}$ defines the statistic, the inequality is componentwise, and $\widetilde{\vec{a}} \overset{d}{=} \vec{a}$; i.e., $\vec{a}$ is the observed treatment assignment vector while $\widetilde{\vec{a}}$ denotes a random realization.
For deterministic $\Mtheta$, we have $\vec{y_{\widetilde{a}}} = \vec{m}_{\vec{\theta}}(\vec{y_a}, \vec{a}, \widetilde{\vec{a}})$.
Otherwise, $\vec{y_{\widetilde{a}}}$ is a stochastic imputation, and the likelihood function includes randomness due to \emph{both} this imputation and the random treatment assignment, $\widetilde{\vec{a}}$.
The density $p(\vec{s} | \vec{\theta}, \vec{y_a})$ is the Radon--Nikodym derivative of $\PP(\vec{s} | \vec{\theta}, \vec{y_a})$ with respect to some dominating measure.
In practice, the likelihood function may be intractable, in which case we can approximate it via asymptotic expressions or simulation-based methods \citep{gutmann_bayesian_2016, li_general_2017}.
Bijective treatment effect models may sometimes result in flat, uninformative likelihood functions.
Section \ref{sec:discrete-stats} discusses this issue and proposes a simple strategy for addressing it.

The prior density $p(\vec{\theta} | \vec{y_a})$ also merits further discussion.
In the regression setting, some authors emphasize that $p(\vec{\beta} | \mat{X})$ reduces to $p(\vec{\beta})$ under certain specifications of the prior distribution \citep[p. 354]{gelman_bayesian_2014}.
Alternatively, we may choose to specify $p(\vec{\beta} | \mat{X})$ directly, potentially using $\mat{X}$ to inform this prior distribution---a classic example being Zellner's $g$-prior \citep{zellner_assessing_1986}.
The BRI framework follows the latter strategy, setting $p(\vec{\theta} | \vec{y_a})$ directly, thereby circumventing the need to specify marginal distributions for the potential outcomes.
This approach results in a simple, robust analysis in which analysts focus their modeling efforts on the causal effects of interest---not error distributions or other potentially high-dimensional nuisance parameters.

\begin{figure}[htbp]
\centering
\begin{subfigure}{0.24\textwidth}
\includegraphics[width=\linewidth]{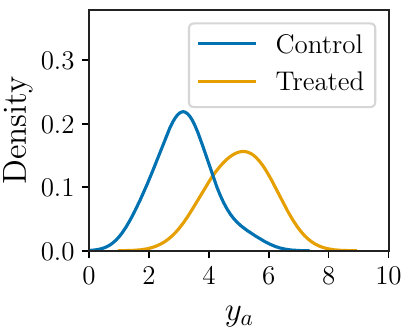}
\caption{Observed Data}
\label{fig:ya-observed}
\end{subfigure}
\begin{subfigure}{0.24\textwidth}
\includegraphics[width=\linewidth]{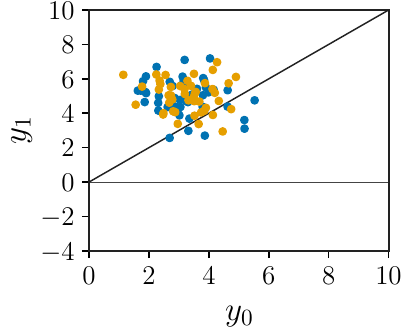}
\caption{Positive Effects}
\label{fig:ya-positive}
\end{subfigure}
\begin{subfigure}{0.24\textwidth}
\includegraphics[width=\linewidth]{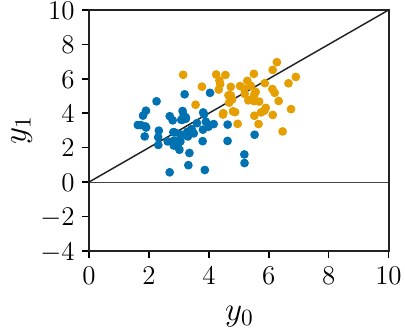}
\caption{Near-Zero Effects}
\label{fig:ya-zero}
\end{subfigure}
\begin{subfigure}{0.24\textwidth}
\includegraphics[width=\linewidth]{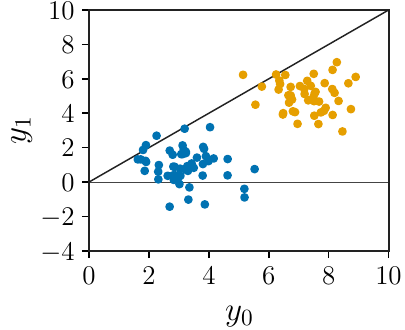}
\caption{Negative Effects}
\label{fig:ya-negative}
\end{subfigure}
\caption{Panel (a) shows KDEs of the observed values of $\vec{y_a}$ for a simulated data set, segmented by treatment assignment. Panels (b)--(d) show three data sets that are consistent with data shown in Panel (a) but which have positive, near-zero, and negative treatment effects, respectively.
}
\label{fig:ya-uninformative}
\end{figure}

Figure \ref{fig:ya-uninformative} illustrates another justification for setting $p(\vec{\theta} | \vec{y_a})$ directly.
Panel \ref{fig:ya-observed} shows kernel density estimates (KDEs) for the observed entries in $\vec{y_0}$ and $\vec{y_1}$ for a simulated data set.
Panels \ref{fig:ya-positive}--\ref{fig:ya-negative} plot complete data sets that are consistent with Panel \ref{fig:ya-observed} but which include vastly different causal effects.
Because the analyst's belief is that $\sigma(\vec{y_a}) \subset \sigma(\mat{Y})$ (see Section \ref{sec:setup}), observing $\vec{y_a}$ provides no information on $\vec{a}$.
Then, supposing the values of $a_i$ are independent, we can produce assignment and counterfactual vectors consistent with Panel \ref{fig:ya-observed} that yield arbitrary causal effects; thus, without imposing additional modeling assumptions, $\vec{y_a}$ provides effectively no information regarding $\vec{\theta}$.
For this reason, we suggest setting $p(\vec{\theta} | \vec{y_a})$ directly based on prior beliefs.
An alternative justification involves decomposing $p(\vec{\theta} | \vec{y_a}) \propto p(\vec{\theta}) p(\vec{y_a} | \vec{\theta})$ and specifying an uninformative likelihood for $\vec{y_a}$.

Online Appendix \ref{app:extensions} details various extensions of the basic BRI framework, including how to introduce covariates, perform sensitivity analysis, jointly estimate assignment mechanisms, and generalize beyond binary treatments.


\subsection{Estimation}
\label{sec:est-inf}

In principle, we can apply any standard Bayesian computational method to estimate BRI models.
The primary challenge compared to standard Bayesian models is that the likelihood typically does not admit a simple closed-form expression.
BRI models can be implemented in modern probabilistic programming languages due to their flexible and extensible interface; our case study in Section \ref{sec:case-study} uses NumPyro \citep{bingham_pyro_2019}.

\section{Discrete Statistics}
\label{sec:discrete-stats}

This section addresses a methodological challenge that can arise with discrete statistics---namely, that the likelihood function can be flat as a function of $\vec{\theta}$.

\subsection{Strategies for Discrete Statistics}
\label{sec:strategies-discrete}

The canonical example in which this challenge arises is the constant treatment effect model with the DIM statistic, $s_{\Delta}$.
Under simple randomization, the likelihood consists of up to $2^n$ atoms.
In fact, if $\vec{y_a}$ is sampled from a distribution that is absolutely continuous with respect to Lebesgue measure, $p(\theta | \vec{y_a})$ is similarly continuous, and $\Pr(a_i = 1) = 1/2$ independently, then $p(s_{\Delta} | \theta, \vec{y_a}) = 1/2^n$ for almost all $\theta$, resulting in an uninformative likelihood function.

Fortunately, several potential resolutions are available.
We may modify the statistic, specifying a statistic that is naturally discrete, such as the Wilcoxon rank-sum (RS) statistic \citep{c4091bd3-d888-3152-8886-c284bf66a93a}.
Alternatively, we could apply asymptotic approximations, employ a stochastic treatment effect model, or ``coarsen'' the observed event.
The latter involves computing the posterior conditional on $\vec{s}$ being in some neighborhood of its observed value.
This event-coarsening strategy is similar to the method proposed in \citet{miller_robust_2019} and the framework of approximate Bayesian computation \citep{beaumont_approximate_2019}.
Our theoretical results in Section \ref{sec:theory} employ this strategy.

\subsection{Discrete Statistic Simulation Study}
\label{sec:discrete-simulation}

This section presents simulation results showing that the above solutions perform adequately and often produce similar inferences.
In the simulation, we assume the constant treatment effect model, $y_{1i} = y_{0i} + \theta$, and sample $\theta \sim \Normaldist(0, 10^2)$.
We draw $\vec{a}$ according to complete randomization with $n_0 \coloneqq \sum_{i=1}^n (1 - a_i) = \sum_{i=1}^n a_i \eqqcolon n_1$.
In the first simulation, we set $n_0 = n_1 = 5$ and compare the following six methods:
\begin{itemize}
    \item Prior: Generate inferences directly from the prior distribution.
    \item DIM: DIM estimator ($s_{\Delta}$) with asymptotically conservative variance estimator.
    \item LIB: An LIB approach using the sampling distribution of $s_{\Delta}$ as the likelihood function.
    \item BRI-A: \textbf{A}symptotic Gaussian approximation to BRI likelihood function with $s_{\Delta}$.
    \item BRI-C: BRI model with $s_{\Delta}$ and the \textbf{C}oarsening strategy.
    \item BRI-RS: BRI model with the Wilcoxon \textbf{RS} statistic.
\end{itemize}

For the Prior method, 95\% credible intervals (CIs) are derived directly from the true data-generating prior for $\theta$.
For the other Bayesian methods, which also use this true prior, we assess whether their 95\% CIs have the appropriate coverage level.
We approximate the posterior distributions on a fine grid from -50 to +50 and perform 10,000 independent repetitions on a personal computer with 48 GB of RAM and 14 CPUs.

\begin{table}[hbtp]
\centering
\begin{tabular}{lrrrr}
\toprule
\textbf{Metrics} & \multicolumn{1}{c}{\textbf{Bias}} & \multicolumn{1}{c}{\textbf{MSE}} & \multicolumn{1}{c}{\textbf{Coverage}} & \multicolumn{1}{c}{\textbf{CI Length}} \\
\midrule
Prior & 0.060 (0.100) & 100.802 (1.431) & 0.950 (0.002) & 39.210 (0.000) \\
DIM & -0.063 (0.071) & 49.968 (0.710) & 0.882 (0.003) & 24.104 (0.062) \\
LIB & -0.031 (0.059) & 35.338 (0.504) & 0.893 (0.003) & 20.155 (0.038) \\
BRI-A & -0.013 (0.062) & 38.725 (0.613) & 0.959 (0.002) & 26.763 (0.046) \\
BRI-C & 0.002 (0.070) & 48.488 (0.847) & 0.967 (0.002) & 30.385 (0.051) \\
BRI-RS & -0.028 (0.063) & 39.589 (0.580) & 0.980 (0.001) & 30.668 (0.043) \\
\bottomrule
\end{tabular}
\caption{\label{tab:discrete-exact} Empirical bias, MSE, 95\% CI coverage, and average 95\% CI length of the six methods in the first simulation study. The values in parentheses denote estimated Monte Carlo standard errors. The BRI methods produce accurate point estimates and approximately calibrated (or conservative) CIs.}
\end{table}

Table \ref{tab:discrete-exact} shows the results from this first simulation.
All of the methods exhibit minimal bias.
The Bayesian methods perform noticeably better than the DIM method in terms of mean squared error (MSE) due to inclusion of the (correct) prior distribution.
The DIM and LIB methods fail to cover at the 95\% level, attaining only 88.2\% and 89.3\% coverage rates, respectively.
In contrast, the BRI models all cover slightly above their nominal level---an artifact of the decoupling of the assignment vector as described in Section \ref{sec:setup}.
In online Appendix \ref{app:simulation-details}, we show that this phenomenon does not occur with oracle methods that observe a new sampled value of $\vec{s}$ based on an independent draw of $\vec{a}$.
BRI can be regarded as an approximation to these exact, oracle methods that are not computable in practice.
Our theoretical results and second simulation study demonstrate that this phenomenon dissipates in large samples, resulting in calibrated coverage rates under correct model specification.

\begin{table}[phtb]
\centering
\setlength{\tabcolsep}{4.5pt}
\begin{tabular}{lrrrrr}
\toprule
\textbf{Metrics} & \multicolumn{1}{c}{$\mathbf{n_0 = n_1}$} & \multicolumn{1}{c}{\textbf{Prior}} & \multicolumn{1}{c}{\textbf{DIM}} & \multicolumn{1}{c}{\textbf{LIB}} & \multicolumn{1}{c}{\textbf{BRI-A}} \\
\midrule
\multirow[c]{4}{4.6em}{\textbf{Bias}} & 10 & 0.042 (0.099) & -0.015 (0.050) & -0.003 (0.045) & 0.006 (0.046)\\
 & 40 & 0.034 (0.100) & -0.010 (0.025) & -0.007 (0.024) & -0.007 (0.024)\\
 & 200 & 0.080 (0.099) & 0.012 (0.011) & 0.013 (0.011) & 0.013 (0.011)\\
 & 1000 & -0.008 (0.100) & -0.004 (0.005) & -0.004 (0.005) & -0.004 (0.005)\vspace{8pt}\\
\multirow[c]{4}{4.6em}{\textbf{MSE}} & 10 & 98.445 (1.385) & 24.903 (0.348) & 20.254 (0.286) & 20.755 (0.297)\\
 & 40 & 99.736 (1.419) & 6.315 (0.088) & 5.935 (0.083) & 5.935 (0.083)\\
 & 200 & 98.272 (1.395) & 1.248 (0.018) & 1.233 (0.017) & 1.233 (0.017)\\
 & 1000 & 99.865 (1.403) & 0.251 (0.004) & 0.250 (0.004) & 0.250 (0.004)\vspace{8pt}\\
\multirow[c]{4}{4.6em}{\textbf{95\% CI Coverage}} & 10 & 0.951 (0.002) & 0.922 (0.003) & 0.925 (0.003) & 0.960 (0.002)\\
 & 40 & 0.950 (0.002) & 0.943 (0.002) & 0.945 (0.002) & 0.955 (0.002)\\
 & 200 & 0.951 (0.002) & 0.948 (0.002) & 0.948 (0.002) & 0.950 (0.002)\\
 & 1000 & 0.952 (0.002) & 0.947 (0.002) & 0.949 (0.002) & 0.949 (0.002)\vspace{8pt}\\
\multirow[c]{4}{4.6em}{\textbf{95\% CI Length}} & 10 & 39.210 (0.000) & 18.328 (0.031) & 16.508 (0.023) & 19.585 (0.029)\\
 & 40 & 39.210 (0.000) & 9.642 (0.008) & 9.368 (0.007) & 9.772 (0.007)\\
 & 200 & 39.210 (0.000) & 4.366 (0.002) & 4.349 (0.002) & 4.387 (0.002)\\
 & 1000 & 39.210 (0.000) & 1.959 (0.000) & 1.967 (0.000) & 1.970 (0.000)\\
\bottomrule
\end{tabular}
\caption{Empirical bias, MSE, 95\% CI coverage, and average 95\% CI length for the four methods in the second simulation study at varying sample sizes. The values in parentheses denote estimated Monte Carlo standard errors. Aside from the naive Prior method, only BRI-A attains near-nominal confidence interval coverage at all values of $n_0$, $n_1$.}
\label{tab:discrete-simulation-asymp}
\end{table}

Table \ref{tab:discrete-simulation-asymp} shows results from the second simulation study, varying $n_0 = n_1 \in \{10, 40, 200, 1000\}$.
Due to computational constraints, we removed the exact BRI methods (BRI-C and BRI-RS), opting to focus on the scalable BRI-A method.
The methods exhibit minimal bias at all sample sizes.
The DIM, LIB, and BRI-A methods produce increasingly similar estimates and inferences as the sample size increases.
The LIB and BRI-A methods achieve noticeably lower MSE than the DIM method with small sample sizes (10, 40) due to inclusion of the prior.
Among DIM, LIB, and BRI-A, only BRI-A achieves near-nominal coverage rates at all sample sizes.
Online Appendix \ref{app:simulation-details} provides additional results and further details on the simulation setup.
For complex models with more parameters, we expect that BRI's strong relative performance would persist for larger sample sizes because the other methods would require relatively more data for their asymptotic approximations to perform well.
BRI also offers the benefit of automatic inference; in contrast, LIB and frequentist approaches often require specialized theory for new problem settings.

\section{Theoretical Results}
\label{sec:theory}

This section develops the asymptotic properties of a large class of parametric BRI models obeying certain regularity conditions.
We prove a Bernstein--von Mises Theorem under potential misspecification of $\Mtheta$, and we use it to derive the asymptotic properties of certain posterior moments.
We provide the proofs in online Appendix \ref{app:proofs}.

\subsection{Theoretical Setup \& Assumptions}
\label{sec:theory-setup}

We now consider a triangular array of random variables with each row equal to $(\vec{a}_n, \vec{y}_{\vec{a}n})$.
We do not impose any parametric distributional assumptions on $\vec{y}_{\vec{a}n}$.
However, we do assume that $\vec{a}_n \sim \PP_n(\vec{a})$ with $\PP_n(\cdot)$ representing the known distribution of $\vec{a}_n$.
We denote the statistic as $\vec{s}_n \coloneqq \vec{f}_n(\vec{y}_{\vec{a}n}, \vec{a}_n)$ and restrict attention to the case $p = k$.
The theoretical results require the following assumptions.
\begin{assumption}
\label{assumption:moments-exist}
The model-based conditional moments $\vec{r}_n(\vec{\theta}) \coloneqq \E \left(\vec{s}_n | \vec{\theta}, \vec{y}_{\vec{a}n}\right)$ and $\mat{V}_n(\vec{\theta}) \coloneqq n \cdot \Var(\vec{s}_n | \vec{\theta}, \vec{y}_{\vec{a}n})$ exist for all $n \in \N$ and $\vec{\theta} \in \mathcal{T}$.
\end{assumption}

The functions $\vec{r}_n$ and $\mat{V}_n$ represent the mean and variance (respectively) of the randomization distribution under the treatment effect model, both of which may depend on $\vec{y}_{\vec{a}n}$.
Although these functions depend on the assumed treatment effect model, most of the theoretical results do not require them to be correctly specified for the nonparametrically defined quantities $\E \left(\vec{s}_n | \vec{y}_{\vec{a}n}\right)$ and $n \cdot \Var(\vec{s}_n | \vec{y}_{\vec{a}n})$.

\begin{assumption}
\label{assumption:compact} 
The parameter space, $\mathcal{T}$, is compact: $\vec{\theta} \in \mathcal{T} \coloneqq \supp(\vec{\theta}) \subset \R^p$.
\end{assumption}

\begin{assumption}
\label{assumption:r}
There exists a function $\vec{r}: \mathcal{T} \to \R^p$ such that
\begin{enumerate}[label=(\alph*)]
    \item $\sqrt{n} \cdot \sup_{\vec{\theta} \in \mathcal{T}} \| \vec{r}_n(\vec{\theta}) - \vec{r}(\vec{\theta}) \|_{\infty} \overset{p}{\to} 0$,
    \item there exists a unique value $\vec{\theta}^* \in \mathcal{T}$ such that $\vec{s}_n \overset{p}{\to} \vec{r}(\vec{\theta}^*)$,
    \item $\vec{r}(\vec{\theta})$ is twice differentiable in an open neighborhood of $\vec{\theta}^*$,
    \item $\vec{r}'(\vec{\theta}^*)$ is invertible, and
    \item $\{\vec{r}'(\vec{\theta^*})\}^{-1} \mat{V}(\vec{\theta^*}) \{\vec{r}'(\vec{\theta^*})\}^{-\top}$ is positive definite.
\end{enumerate}
\end{assumption}

Although $\vec{r}_n$ is (in general) random, Assumption \ref{assumption:r} requires it to converge to a twice-differentiable function at a rate faster than $\sqrt{n}$.
Condition (b) is required for unique identification of $\vec{\theta}$.
Conditions (d) and (e) ensure a non-degenerate limiting distribution.

\begin{assumption}
\label{assumption:V}
There exists a function $\mat{V}: \mathcal{T} \to \R^{p \times p}$ such that
\begin{enumerate}[label=(\alph*)]
    \item $\sup_{\vec{\theta} \in \mathcal{T} }\|\mat{V}_n(\vec{\theta}) - \mat{V}(\vec{\theta})\|_{\infty,\infty} \overset{p}{\to} 0$,
    \item for all $\vec{\theta} \in \mathcal{T}$, $\mat{V}(\vec{\theta})$ is bounded as $\mat{V}_{\min} \preceq \mat{V}(\vec{\theta}) \preceq \mat{V}_{\max}$ for two positive definite matrices $\mat{V}_{\min}, \mat{V}_{\max} \in \R^{p \times p}$, and
    \item $\mat{V}(\vec{\theta})$ is continuous in an open neighborhood of $\vec{\theta}^*$.
\end{enumerate}
\end{assumption}

Assumption \ref{assumption:V} ensures that the variance of the randomization distribution converges appropriately across all values of $\vec{\theta}$.

\begin{assumption}
\label{assumption:prior}
The prior distribution $p(\vec{\theta} | \vec{y}_{\vec{a}n})$ satisfies the following conditions:
\begin{enumerate}[label=(\alph*)]
    \item there exists $C > 0$ such that $1\{\sup_{\vec{\theta} \in \mathcal{T}} p(\vec{\theta} | \vec{y}_{\vec{a}n}) \leq C\} \overset{p}{\to} 1$ as $n \to \infty$,
    \item there exists $C > 0$ such that $p\left(\vec{\theta}^* \big| \vec{y}_{\vec{a}n}\right) \overset{p}{\to} C$ as $n \to \infty$, and
    \item for any $\epsilon > 0$, there exists $\delta > 0$ such that
    \[
    1\left\{\sup_{\vec{\theta} \in \R^p: \|\vec{\theta} - \vec{\theta}^*\|_{\infty} < \delta}\left|p\left(\vec{\theta} \big| \vec{y}_{\vec{a}n}\right) - p\left(\vec{\theta}^* \big| \vec{y}_{\vec{a}n}\right) \right| < \epsilon\right\} \overset{p}{\to} 1\text{ as } n \to \infty.
    \]
\end{enumerate}
\end{assumption}

Assumption \ref{assumption:prior} requires that (a) the prior is uniformly bounded with high probability, (b) the prior density at $\vec{\theta}^*$ converges in probability to a positive constant, and (c) the prior density near $\vec{\theta}^*$ is close to the prior density at $\vec{\theta}^*$ with high probability for large $n$.
These conditions allow $p(\vec{\theta} | \vec{y}_{\vec{a}n})$ to depend on $\vec{y}_{\vec{a}n}$ without weakening the theoretical results.

Below, we denote the probability density and cumulative distribution functions of the multivariate Gaussian distribution with parameters $\vec{\mu}, \mat{\Sigma}$ as $\phi(\cdot; \vec{\mu}, \mat{\Sigma})$ and $\Phi(\cdot; \vec{\mu}, \mat{\Sigma})$, respectively.

\begin{assumption}
\label{assumption:clt}
    Let $\vec{z}_n(\vec{\theta}) \coloneqq \sqrt{n}\{\vec{s}_n - \vec{r}_n(\vec{\theta})\}$ and $\delta \in (0, 1)$.
    Then, under correct specification of the treatment effect model, there exist $C \in \R$ and $N \in \N$ such that \[\sup_{\vec{\theta} \in \mathcal{T}} \left\vert \Pr\left\{\vec{z}_n(\vec{\theta}) \leq \vec{z} | \vec{\theta}, \vec{y}_{\vec{a}n}\right\} - \Phi\left\{\vec{z; \vec{0}, \mat{V}_n(\vec{\theta})}\right\} \right\vert \leq C/\sqrt{n}\]
    with probability at least $\delta$ for all $n \geq N$ and $\vec{z} \in \R^p$.
\end{assumption}

Assumption \ref{assumption:clt} requires a Central Limit Theorem to hold for the randomization distribution with a corresponding Berry--Esseen bound.
Classical Berry--Esseen bounds rely on independent observations;
however, there are extensions to combinatorial CLTs \citep{shi_Berry--Esseen_2023}.
Most of these bounds involve the third absolute moment, and some also require that the fourth moment is bounded.
The difference between Assumption \ref{assumption:clt} and the results cited above is that Assumption \ref{assumption:clt} requires uniformity over $\vec{\theta}$.
Because the randomization distribution involves model-adjusted potential outcomes, the regularity conditions for the CLTs cited above require moments of the imputed potential outcomes to be bounded \emph{uniformly} over $\vec{\theta}$.
Due to Assumption \ref{assumption:compact}, this requirement is satisfied under the constant treatment effect model provided the moment conditions hold on the (original) potential outcomes.

Although we state Assumptions \ref{assumption:r}--\ref{assumption:clt} and the theoretical results in terms of convergence in probability, similar results can be obtained in terms of almost-sure convergence under slightly stronger assumptions.

\subsection{Bernstein--von Mises Theorem}
\label{sec:bvm}

Bernstein--von Mises Theorems provide conditions under which a Bayesian posterior distribution is well approximated by a limiting Gaussian distribution.
The classical Bernstein--von Mises Theorem applies to independent and identically distributed data sampled from a parametric model.
The theorem has since been extended to misspecified models \citep{kleijn_bernstein-von-mises_2012}, semiparametric and nonparametric models \citep{bickel_semiparametric_2012, rousseau_frequentist_2016}, and generalized posterior distributions \citep{miller_asymptotic_2021}.

We employ event coarsening (see Section \ref{sec:strategies-discrete}) to avoid flat likelihoods for discrete $\vec{s}_n$.
Specifically, we consider neighborhoods of the form $\mathcal{N}_{\epsilon_n}(\vec{s}_n^*) \coloneqq\{\vec{s}_n \in \R^p: \|\vec{s}_n - \vec{s}_n^*\|_{\infty} \leq \epsilon_n\}$, where $\vec{s}_n^*$ is the observed statistic value and $\epsilon_n = o(n^{-1/2})$.
Under this choice of neighborhood, we can approximate the likelihood function as follows.

\begin{lemma}
\label{lemma:likelihood}
Let $\delta \in (0, 1)$, $\alpha \in (0.5, \frac{p+1}{2p})$, $\epsilon_n = n^{-\alpha}$, and $\gamma \coloneqq \max\{p(\alpha - 0.5) - 0.5, 0.5 - \alpha\} < 0$. Then, under Assumptions \ref{assumption:consistency}--\ref{assumption:clt}, there exists $C \in \R$ and $N \in \N$ such that
\begin{equation}
\label{eq:likelihood-approx}
\left|
\frac{\Pr\left\{\vec{s}_n \in \mathcal{N}_{\epsilon_n}(\vec{s}_n^*)| \vec{\theta}, \vec{y}_{\vec{a}n}\right\}}{(2\epsilon_n \sqrt{n})^p}
-
\phi\left[\sqrt{n}\left\{\vec{s}_n^* - \vec{r}_n(\vec{\theta})\right\}, \vec{0}, \mat{V}_n(\vec{\theta})\right]
\right|
\leq
C n^{\gamma}
\end{equation}
with probability at least $1 - \delta$ for all $n \geq N$ and $\vec{\theta} \in \mathcal{T}$.
This bound is optimized with
\[
\alpha = \frac{2+p}{2(p+1)},\quad \gamma = -\frac{1}{2(p+1)}.
\]
\end{lemma}
Using Lemma \ref{lemma:likelihood}, we can prove the following Bernstein--von Mises theorem.
\begin{theorem}[Bernstein--von Mises]
\label{thm:bvm}
Under the Assumptions of Lemma \ref{lemma:likelihood}, the posterior distribution converges in total variation distance as follows:
\[
\int_{\vec{\theta \in \mathcal{T}}}
\left|
    p\left\{\vec{\theta} | \vec{y}_{\vec{a}n}, \vec{s}_n \in \mathcal{N}_{\epsilon_n}\left(\vec{s}_n^*\right)\right\} -
    \phi\left(
        \vec{\theta},
        \vec{\mu}_n,
        \mat{\Sigma}/n
    \right)
\right| d\vec{\theta} \overset{p}{\to} 0
\]
as $n \to \infty$, where $\vec{\mu}_n \coloneqq \vec{\theta}^* + \{\vec{r}'(\vec{\theta^*})\}^{-1} \{\vec{s}_n - \vec{r}(\vec{\theta^*})\}$, $\mat{\Sigma} \coloneqq \{\vec{r}'(\vec{\theta^*})\}^{-1} \mat{V}(\vec{\theta^*}) \{\vec{r}'(\vec{\theta^*})\}^{-\top}$, and $\vec{s}_n^*$ is the observed value of $\vec{s}_n$.
\end{theorem}

Theorem \ref{thm:bvm} guarantees that the posterior distribution is approximately Gaussian in large samples.
Informally, the theorem states that $\sqrt{n}(\vec{\theta} - \vec{\mu}_n) \overset{d}{\to} \Normaldist(\vec{0}, \mat{\Sigma})$ with the left-hand side representing the posterior distribution (i.e., $\vec{\theta}$ is viewed as random with $\vec{\mu}_n$ fixed).

Theorem \ref{thm:bvm} also provides insight into the behavior of BRI under potential misspecification of the treatment effect model.
Specifically, it guarantees that the posterior distribution will concentrate around $\vec{\theta}^*$: the value of $\vec{\theta}$ such that the limits of $\vec{s}_n$ and $\vec{r}_n(\vec{\theta})$ coincide.
Under misspecification, $\vec{\theta}^*$ need not correspond with a model parameter; however, we show in Section \ref{sec:theory-example} that it is sometimes possible to specify BRI models for which $\vec{\theta}^*$ is an interpretable causal quantity, such as an average treatment effect (ATE).
Theorem \ref{thm:bvm} further enables us to determine the frequency properties of certain posterior functionals.
\begin{corollary}
\label{corollary:posterior-mean}
Let $\widehat{\vec{\theta}}_n$ denote the BRI posterior mean, defined as
\[
\widehat{\vec{\theta}}_n \coloneqq
\frac
    {\int_{\vec{\theta} \in \mathcal{T}} \vec{\theta} \cdot  p(\vec{\theta} | \vec{y}_{\vec{a}n}) p(\vec{s}_n | \vec{\theta}, \vec{y}_{\vec{a}n}) d\vec{\theta}}
    {\int_{\vec{\theta} \in \mathcal{T}} p(\vec{\theta} | \vec{y}_{\vec{a}n}) p(\vec{s}_n | \vec{\theta}, \vec{y}_{\vec{a}n}) d\vec{\theta}}.
\]
Then $\widehat{\vec{\theta}}_n$ is asymptotically equivalent to $\vec{\mu}_n$ in the sense that $\sqrt{n}\|\widehat{\vec{\theta}}_n - \vec{\mu}_n \|_{\infty} \overset{p}{\to} 0$.
Moreover, for the posterior covariance matrix $\widehat{\mat{\Sigma}}_n$ (defined similarly), we also have $n \cdot \widehat{\mat{\Sigma}}_n \overset{p}{\to} \mat{\Sigma}$.
\end{corollary}
Corollary \ref{corollary:posterior-mean} is stronger than the assertion that $\widehat{\vec{\theta}}_n \overset{p}{\to} \vec{\theta}^*$ and $\vec{\mu}_n \overset{p}{\to} \vec{\theta}^*$.
It implies that $\widehat{\vec{\theta}}_n$ and $\vec{\mu}_n$ produce the same asymptotic inferences, so we can determine the frequency properties of $\widehat{\vec{\theta}}_n$ from $\vec{\mu}_n$.
In fact, we can show asymptotic equivalence between $\vec{\mu}_n$ and a class of estimators known as Hodges--Lehmann estimators, which are formed by equating statistics with their expectations under a sequence of null distributions \citep{hodges_estimates_1963, rosenbaum_hodges-lehmann_1993, rosenbaum_observational_2002}.
In our notation, these estimators are formed by equating $\vec{s}_n$ with $\vec{r}_n(\vec{\theta})$ and solving for $\vec{\theta}$, which yields $\widetilde{\vec{\theta}}_{n} = \vec{\theta}^* + \{\vec{r}'(\vec{\theta}^*)\}^{-1}\{\vec{s}_n - \vec{r}(\vec{\theta}^*)\} + o_p(n^{-1/2})$.
\begin{corollary}
\label{corollary:hodges-lehmann}
Let $\widetilde{\vec{\theta}}_n$ denote the Hodges--Lehmann estimator.
Then the estimators $\widetilde{\vec{\theta}}_n$, $\widehat{\vec{\theta}}_n$, and $\vec{\mu}_n$ are asymptotically equivalent in the sense that
\[\sqrt{n}\max\left(
\|\widetilde{\vec{\theta}}_n - \widehat{\vec{\theta}}_n \|_{\infty},
\|\widetilde{\vec{\theta}}_n - \vec{\mu}_n \|_{\infty},
\|\widehat{\vec{\theta}}_n - \vec{\mu}_n \|_{\infty}
\right)\overset{p}{\to} 0.\]
\end{corollary}
To further explore the implications of Theorem \ref{thm:bvm}, we derive the frequentist properties of $\vec{\mu}_n$ under correct model specification.
\begin{theorem}
\label{thm:mu_n}
Under correct specification of the treatment effect model with model parameter $\vec{\theta}^*$,  we have
\[
\sqrt{n}\mat{\Sigma}^{-1/2}(\vec{\mu}_n - \vec{\theta}^*)
\overset{d}{\to}
\Normaldist(\vec{0}, \mat{I}_p).
\]
\end{theorem}
Because the asymptotic sampling distribution of $\vec{\mu}_n$ in Theorem \ref{thm:mu_n} corresponds with the asymptotic posterior distribution in Theorem \ref{thm:bvm}, under correct model specification BRI will deliver valid frequentist inferences provided they are based on sufficiently well-behaved posterior functionals, such as posterior moments or quantiles; see \citet[Section 10.3]{vaart_asymptotic_1998} or \citet{bochkina_bernsteinvon_2014}.

More generally, when the treatment effect model is misspecified, Theorem \ref{thm:mu_n} does not apply.
In particular, Assumption \ref{assumption:clt} does not necessarily guarantee asymptotic normality of $\vec{\mu}_n$, and its moments may not equal those of Theorem \ref{thm:mu_n}.
Instead, they are given by
\begin{align*}
\E(\vec{\mu}_n | \vec{y}_{\vec{a}n}) &=
\vec{\theta}^* + \{\vec{r}'(\vec{\theta^*})\}^{-1} \{\E(\vec{s}_n | \vec{y}_{\vec{a}n}) - \vec{r}(\vec{\theta^*})\},\\
\Var(\vec{\mu}_n | \vec{y}_{\vec{a}n}) &=
\{\vec{r}'(\vec{\theta^*})\}^{-1} \Var(\vec{s}_n | \vec{y}_{\vec{a}n}) \{\vec{r}'(\vec{\theta^*})\}^{-\top},
\end{align*}
provided they exist.
In this case, the posterior mean may still be viewed as an estimate of $\E(\vec{\mu}_n | \vec{y}_{\vec{a}n})$.
However, $\Var(\vec{s}_n | \vec{y}_{\vec{a}n})$ will not necessarily equal $\mat{V}(\vec{\theta^*})$, leading to incorrect uncertainty quantification even for large $n$.
The plug-in approach proposed in \citet{leavitt_randomization-based_2023} achieves correct asymptotic coverage under misspecification by effectively replacing $\mat{V}(\vec{\theta^*})$ with a conservative variance estimate.
This approach, although not dogmatically Bayesian, could similarly be applied within our framework to obtain asymptotically valid inference even under misspecification.

\subsection{Theory Example}
\label{sec:theory-example}

This section provides a simple example to demonstrate the theoretical results of Section \ref{sec:bvm}.
We assume the constant treatment effect model and complete randomization of $\vec{a}_n$ with fixed treatment proportion $\pi$ (so $n_1 = n \pi$), and we employ the DIM statistic, $s_{\Delta}$.
To simplify the analysis, we consider an asymptotic regime in which $n_1 \coloneqq n \pi \in \N$ and define $n_0 \coloneqq n - n_1$.
We can then show that $r(\theta) = r_n(\theta) = \theta$, so $\theta^* = \E(y_{1i} - y_{0i})$ under a superpopulation assumption;
thus, the BRI posterior will concentrate around the ATE, $\E(y_{1i} - y_{0i})$, even if the constant treatment effect model is misspecified.
We can further show that $\mu_n = \widetilde{\theta}_n = s_{\Delta n}$ so that the posterior mean is asymptotically equivalent to $s_{\Delta n}$.

The model-based variance is $v_n(\theta) = \Varhat(\vec{y}_{\vec{a}n} - \vec{a}_n\theta) / \{\pi (1-\pi)\}$, which gives $v(\theta^*) = \Var(y_{1i})/(1-\pi) + \Var(y_{0i})/\pi$.
In contrast, the (conservative) frequency-based variance is $\{\Var(y_{1i})/\pi + \Var(y_{0i})/(1-\pi)\}/n$.
Thus, under misspecification of the treatment effect model, BRI's asymptotic posterior variance will match the frequentist variance provided $\pi=0.5$ or $\Var(y_{0i}) = \Var(y_{1i})$, the latter being an implication of the constant treatment effect model; these conditions mirror those given in \citet{Romano01091990} and \citet{10.1214/13-AOS1090} under which permutation tests are asymptotically robust.
Online Appendix \ref{app:additional-theory-examples} provides examples in which the posterior mean is asymptotically equivalent to inverse-probability-weighted and H\'ajek estimators.

\section{Case Study}
\label{sec:case-study}

This section provides a re-analysis of two randomized controlled trials in a virtual fast-food restaurant \citep{marty_socioeconomic_2020}.
Because the design of the two trials is the same, we analyze them as a single experiment.
The protocol and data for the original publication are publicly available at \url{https://osf.io/ajcr6/}.

\subsection{Experimental Design \& Data}
\label{sec:exp-design-data}

The experiment includes 1,743 United Kingdom residents 18 years or older with no dietary restrictions.
Participants interacted with a virtual fast-food restaurant environment modeled after a popular fast-food chain, navigating through the restaurant via mouse clicks and placing an order with a virtual cashier.
Participants were independently randomized with equal probability to one of four experimental conditions in a $2 \times 2$ factorial design with the experimental condition determining the structure of the menu boards.
The two experimental factors were (a) availability of low-calorie foods (75\% vs. 25\% options lower energy) and (b) menu energy labeling (present vs. absent).

The primary research outcome for the study is the total number of calories ordered, summing over the main dish, side, and drink.
The researchers also collected a number of baseline covariates, including education level, frequency of fast-food consumption, and various psychological measures.
In the original data analysis, the researchers analyzed the experiment using analysis of covariance (ANCOVA).
As hypothesized, they found a negative and statistically significant effect for availability of lower energy options on average calories ordered (-71 kcal, $p < 0.001$).
In contrast, the observed difference between labeling vs. no labeling was much smaller and not statistically significant (-18 kcal, $p=0.116$).
The researchers did not find significant evidence of effect moderation.

\subsection{Data Analysis}
\label{sec:data-analysis}

In analyzing the data, our primary goal is to demonstrate the BRI analytic process.
We pay particular attention to the issue of model specification, following the model-checking procedures described in online Appendix \ref{app:model-checking}.
To simplify the analysis, we restrict our attention to a single treatment variable: the availability of healthy options.
We let $a_i=0,1$ denote the groups with 25\%, 75\% healthy options, respectively.

\begin{figure}[htbp]
\centering
\begin{subfigure}{0.32\textwidth}
\includegraphics[width=\linewidth]{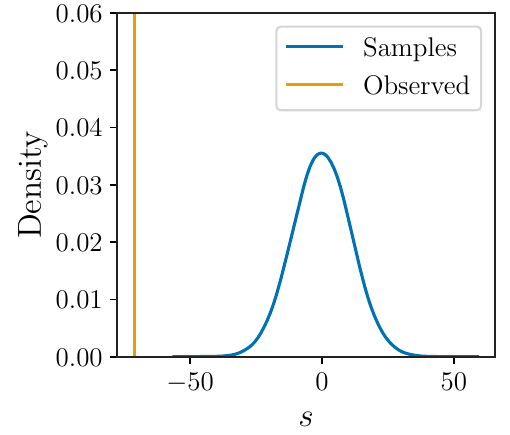}
\caption{FRT}
\label{fig:case-frt}
\end{subfigure}
\begin{subfigure}{0.32\textwidth}
\includegraphics[width=\linewidth]{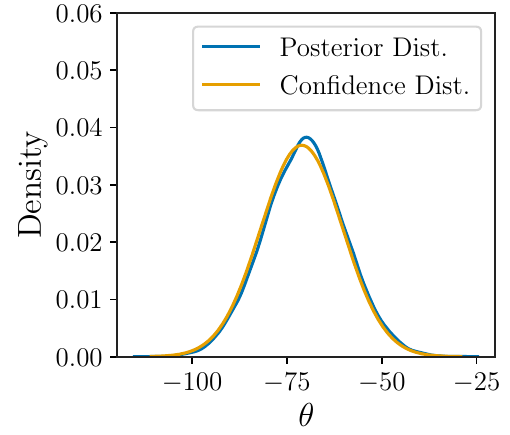}
\caption{Posterior}
\label{fig:case-constant-posterior}
\end{subfigure}
\begin{subfigure}{0.32\textwidth}
\includegraphics[width=\linewidth]{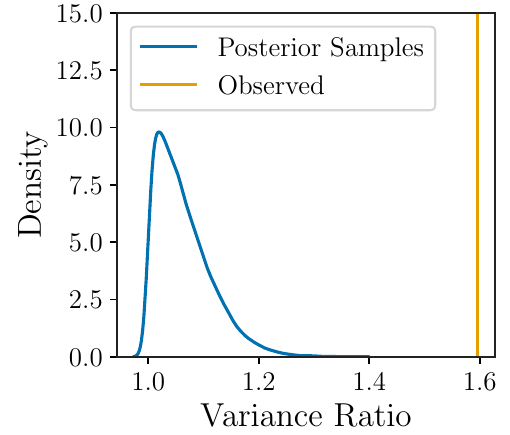}
\caption{Model Check}
\label{fig:var-ratio}
\end{subfigure}
\caption{Panel (a) plots an FRT testing the sharp null hypothesis of no effect, resulting in rejection ($p < 10^{-5}$). Panel (b) compares the posterior distribution of the BRI constant-effect model to its asymptotic approximation from Theorem \ref{thm:bvm}---the frequentist confidence distribution. Panel (c) shows the result of a posterior model check (an embedded FRT), indicating that the constant-effect model does not capture the different group-level variances.}
\label{fig:case-distributions}
\end{figure}

We first consider the constant treatment effect model and DIM statistic, $s_{\Delta}$.
We perform an FRT against the null hypothesis that $\theta$, the assumed-constant effect, is zero.
Figure \ref{fig:case-frt} plots the randomization distribution against the observed statistic value of $-71$.
Of the 100,000 simulated values from the randomization distribution, none exceed 71 in absolute value, resulting in a rejection of the sharp null.
Figure \ref{fig:case-constant-posterior} plots the BRI posterior distribution for the same model with $\theta \sim \Normaldist(0, 100^2)$ compared to its asymptotic approximation from Theorem \ref{thm:bvm}.
Across all models considered, we draw 40,000 posterior samples from a No-U-Turn Sampler (NUTS) using the same computing environment as Section \ref{sec:discrete-simulation}, discarding the first 20,000 as warmup iterations.
For each estimated parameter, we compute the Gelman--Rubin statistic \citep{gelman_inference_1992} by splitting the posterior samples, resulting in a value of $1.00$ in all cases.
The NUTS algorithm fits the constant-effect model in less than four seconds with an analytic large-sample Gaussian approximation of the likelihood function, producing an effective sample size of over 7,000.

\begin{figure}[hbt]
\centering
\begin{subfigure}{0.32\textwidth}
\includegraphics[width=\linewidth]{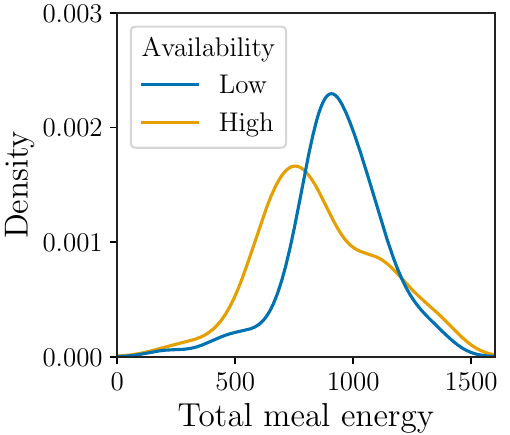}
\caption{Observed Data}
\label{fig:data-kde}
\end{subfigure}
\begin{subfigure}{0.32\textwidth}
\includegraphics[width=\linewidth]{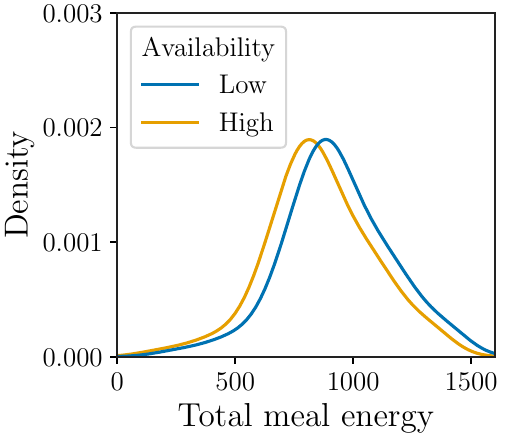}
\caption{Constant Effects}
\label{fig:constant-kde}
\end{subfigure}
\begin{subfigure}{0.32\textwidth}
\includegraphics[width=\linewidth]{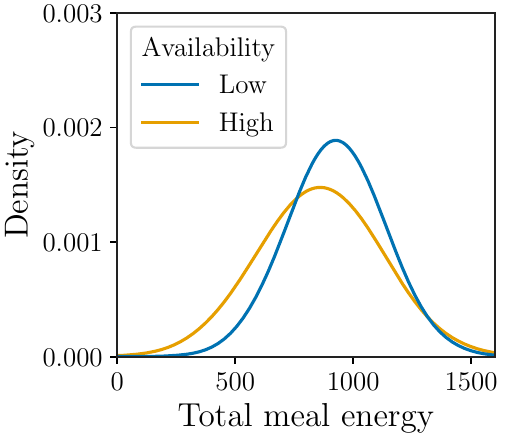}
\caption{2-D Gaussian}
\label{fig:gaussian-kde}
\end{subfigure}
\caption{Panel (a) displays KDEs for the observed data by group. Panel (b) displays KDEs for posterior-averaged imputations from the constant-effects BRI model, highlighting that the constant-effects model fails to capture the heteroscedasticity in the data. Panel (c) plots imputations from a Gaussian superpopulation model; although this model captures the heteroscedasticity, it does not adequately model the higher-order moments in the data, such as the skew.}
\label{fig:constant-checks}
\end{figure}

We perform two types of model checks for this analysis.
First, for each covariate, we check for evidence of moderation by computing the absolute difference in group-wise slopes (via ordinary least squares) and comparing this value to its randomization distribution.
The minimum of the resulting ten $p$-values is 0.09, indicating minimal evidence of treatment effect moderation.
Second, we compute the group-specific sample variances, $\widehat{\sigma}^2_0$ and $\widehat{\sigma}^2_1$, and compare $\max(\widehat{\sigma}^2_0, \widehat{\sigma}^2_1) / \min(\widehat{\sigma}^2_0, \widehat{\sigma}^2_1) \approx 1.6$ to its randomization distribution, averaging over the posterior uncertainty in the model parameter; see Figure \ref{fig:var-ratio}.
In this case, none of the 20,000 samples exceeds 1.6, providing strong evidence against the constant-effect model.
Figures \ref{fig:data-kde} and \ref{fig:constant-kde} illustrate the issue: the constant-effects model fails to capture the increased variance of the High group compared to the Low group.

Figure \ref{fig:gaussian-kde} plots posterior samples from a 2-D Gaussian superpopulation model.
This method addresses the group-level heteroscedasticity, but it does not adequately model the higher-order empirical moments in Figure \ref{fig:data-kde}.
Improving the model fit within the superpopulation framework would require a more flexible model, such as a mixture model, with substantially more parameters.
Below, we show that the BRI analysis can adequately address this challenge with relatively few parameters.

To address the difference in variance between groups, we adopt model \eqref{eq:gaussian-reg}: $y_{1i} | y_{0i} \overset{ind}{\sim} \text{Normal}(\alpha + \beta y_{0i}, \sigma^2)$.
We employ the statistic $(\Bar{y}_1, \widehat{\sigma}_1)$, the sample mean and standard deviation among treated individuals, and assume the prior distribution $\alpha \sim \Normaldist(0, 1,000^2)$, $\beta \sim \Gammadist(1, 1)$, and $\sigma \sim \HalfNormaldist(100^2)$, independently.
We approximate the likelihood function as a Gaussian distribution with mean vector and covariance matrix estimated from 1,000 independent Monte Carlo draws per iteration of the NUTS algorithm---an approach referred to as ``synthetic likelihood'' by \citet{WoodSimonN.2010Sifn} and \citet{gutmann_bayesian_2016};
the algorithm produces over 500 effective samples per parameter in 1.5 hours.

We apply two-sided model checks similar to that of Figure \ref{fig:var-ratio} based on the first five centered and scaled moments of the distribution of $y_{1i}$: $m_1=\Bar{y}_1$, $m_2 = \widehat{\sigma}_1$, and $m_j = \sum_{i=1}^n a_i\{(y_{1i} - \Bar{y}_1)/\widehat{\sigma}_1\}^j /n_1$ for $j=3,4,5$.
The resulting posterior $p$-values are 0.49, 0.50, 0.03, 0.05, and 0.19; thus, this model adequately captures differences between groups in the first and second moments but not the third and fourth.
To better capture these higher-order differences, we fit a final model that allows a more flexible form for $\PP(y_{1i} | y_{0i})$:
\begin{equation}
\label{eq:spline-model}
y_{1i} | y_{0i} \overset{ind}{\sim} \Normaldist\left\{\alpha + g(y_{0i}, \vec{\beta}), \sigma^2\right\},
\quad 
g(y_{0i}, \vec{\beta}) = \int_{0}^{y_{0i}} \exp\left(\sum_{j=0}^3 \beta_j t^j\right) dt.
\end{equation}
Model \eqref{eq:spline-model} ensures that $\E(y_{1i} | y_{0i})$ is an increasing function of $y_{0i}$, a structural constraint that we would expect to hold in this application.
To improve mixing of the NUTS algorithm, we reparameterize model \eqref{eq:spline-model} in terms of standardized outcomes and set the priors as $\sigma \sim \HalfNormaldist(100^2)$, $\alpha \sim \Normaldist(-4, 2^2)$, $\beta_0 \sim \Normaldist(-5.5, 1)$, $\beta_1 \sim \Normaldist(0, 0.5^2)$, $\beta_2 \sim \Normaldist(0, 0.2^2)$, and $\beta_3 \sim\Normaldist(0, 0.1^2)$, independently.
We specify the statistic as $(m_1, m_2, m_3, m_4, m_5)$.
As with model \eqref{eq:gaussian-reg}, we draw 1,000 independent Monte Carlo samples of the statistic per iteration of the NUTS algorithm.
This model fits in 3.4 hours, producing 500--2,200 effective samples per parameter.
The model checks described above result in posterior $p$-values in the range 0.39--0.55, providing little to no evidence against this model in terms of the first five moments of $y_{1i}$.
See Figures \ref{fig:moment-p-values} and \ref{fig:y1} in online Appendix \ref{app:additional-application-results} for visualizations of these posterior $p$-values and the estimated distributions of $y_{1i}$.

\begin{figure}[htbp]
\centering
\begin{subfigure}{0.49\textwidth}
\includegraphics[width=\linewidth]{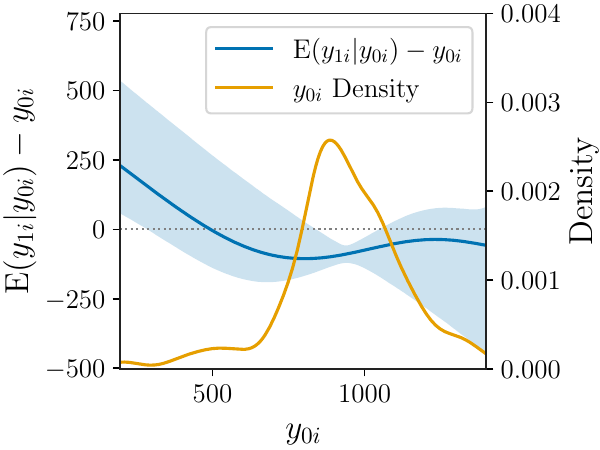}
\caption{$\E(y_{1i} | y_{0i}) - y_{0i}$}
\label{fig:calorie-mean}
\end{subfigure}
\begin{subfigure}{0.49\textwidth}
\includegraphics[width=\linewidth]{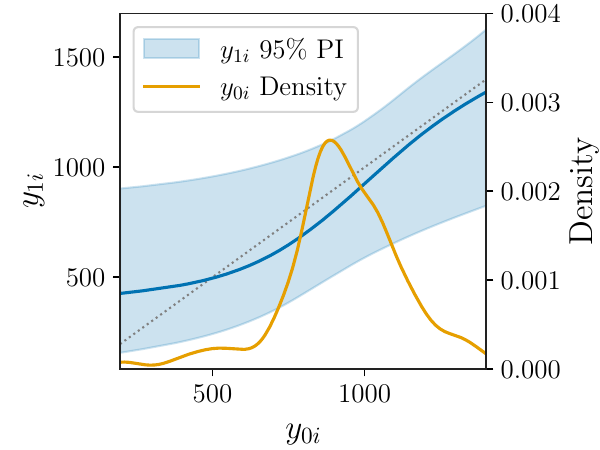}
\caption{$\PP(y_{1i} | y_{0i})$}
\label{fig:calorie-dist}
\end{subfigure}
\caption{Panel (a) plots the posterior distribution of $\E(y_{1i} | y_{0i}) - y_{0i}$ for model \eqref{eq:spline-model}, and Panel (b) plots the posterior predictive distribution $\PP(y_{1i} | y_{0i})$ for the same model. The blue bands indicate 95\% CIs and prediction intervals in Panels (a) and (b), respectively.}
\label{fig:spline-fit}
\end{figure}

Figure \ref{fig:spline-fit} plots the fit for model \eqref{eq:spline-model}.
Figure \ref{fig:calorie-mean} shows that we have significant posterior evidence of a negative effect for $y_{0i}$ in the approximate range $[840, 1050]$.
However, the 95\% credible bands include zero for most other values of $y_{0i}$, indicating that the collected data do not provide strong posterior evidence of an effect for individuals with particularly high ($>1050$) or low ($<840$) values of $y_{0i}$, except perhaps $y_{0i} \leq 300$ for which the estimated effect is positive; though, the latter range includes only ten participants.
Figure \ref{fig:calorie-dist} shows the posterior predictive distribution of $y_{1i} | y_{0i}$.
In this case, the intervals are considerably wider due to the estimated degree of effect heterogeneity at the participant level.

\subsection{Result Summary}
\label{sec:result-summary}

In summary, the BRI modeling process allows for a richer causal analysis compared to classical moment-based estimators, and it adds robustness to the Bayesian approach by removing unnecessary modeling assumptions on the marginal outcome distributions.
For identified parametric models, BRI maintains many of the desirable frequentist properties of robust moment-based estimators, but it also empowers the analyst to explore complex effect heterogeneity with the expressiveness and modularity of Bayesian modeling.

In this case study, BRI uncovers strong evidence of effect heterogeneity, but the heterogeneity is not explained by the observed baseline characteristics.
Instead, our best-fitting model suggests that the effects vary according to $y_{0i}$: the number of calories a participant would order under low availability of healthy options.
We have evidence of a negative effect only for individuals that order a near-average number of calories.
This finding suggests that future work could examine how the effects of structural menu interventions differ based on the number of calories that individuals typically order.
For instance, we might hope to uncover whether these interventions are effective for individuals who most need them: those ordering meals with the highest energy content relative to their caloric needs.
Future work could explore this possibility with more complex experimental designs (e.g., placing participants in multiple conditions sequentially) or by identifying potential moderators more closely related to baseline order size.


\section{Discussion}
\label{sec:discussion}

This article introduces BRI, a framework for robust Bayesian inference of causal effects based principally on the physical act of randomization.
In essence, BRI is a Bayesian analog to randomization-based causal inference methods in that the BRI likelihood function is a randomization distribution of an analyst-specified statistic.
Compared to Bayesian superpopulation models for causal inference with binary treatments, BRI requires weaker assumptions because it treats the observed potential outcomes ($\vec{y_a}$) as fixed quantities, removing the need to specify marginal outcome distributions.
This aspect of BRI enables analysts to focus their modeling efforts on the treatment effects, fitting models that account for complex participant-level effect heterogeneity.

In addition to outlining the basic BRI framework, we also discuss strategies for handling discrete statistics, illustrate how to perform Bayesian model checking via embedded FRTs, and provide theoretical results for a large class of parametric BRI models.
The main result is a Bernstein--von Mises Theorem that guarantees asymptotic Gaussian behavior of the posterior distribution under reasonable assumptions.
We further analyze the asymptotic behavior of the posterior mean, demonstrating asymptotic equivalence with Hodges--Lehmann estimators.
To ensure broad applicability, our theoretical results employ the event-coarsening strategy of Section \ref{sec:strategies-discrete}.
However, this strategy results in an error bound that decays slowly in $n$ for models with moderate to large dimension (see Lemma \ref{lemma:likelihood}).
Future work could develop specialized theory for models where event-coarsening is not needed (such as the BRI-RS model from Section \ref{sec:discrete-simulation}), which we expect would result in a standard $O(n^{-1/2})$ error bound.
Future theoretical work could also develop specialized theory for other specific settings, such as partially identified models or models with jointly estimated assignment mechanisms.
These extensions would require additional assumptions and regularity conditions beyond those given in Section \ref{sec:theory}.

In principle, the BRI framework is applicable to any causal inference problem with a randomized treatment.
Although we emphasize binary treatments with a known assignment mechanism, extensions to many other settings are conceptually straightforward and are outlined in online Appendix \ref{app:extensions}.
Future work could further investigate these extensions, especially the observational setting in which the assignment mechanism must be estimated.
Other interesting extensions include (a) tailored computational approaches for randomization-based likelihood functions and (b) adaptations to more complex settings, such as quasi-experimental designs, dynamic treatment regimes, and instrumental-variable analyses.

We illustrate the BRI analytical process in Section \ref{sec:case-study} in the context of a nutrition experiment that tests structural menu modifications in a virtual restaurant environment.
In this case study, BRI uncovers strong evidence of effect heterogeneity and allows the analyst to fit models to explain it.
Our best-fitting model, a shape-constrained spline-based model, provides strong posterior evidence of a negative treatment effect for individuals who order a near-average (in the range [840, 1050]) number of calories.
However, our estimates for individuals outside this narrow range show much higher posterior uncertainty, indicating limited knowledge of their causal effects.
Thus, the BRI framework enables a richer analysis compared to classical moment-based methods, producing insights and new hypotheses that might otherwise be missed.

\section{Disclosure Statement}\label{sec:disclosure-statement}

The authors report there are no competing interests to declare.

\section{Data Availability Statement}\label{sec:data-availability-statement}

The data for the case study is publicly available through the Open Science Framework at \url{https://osf.io/ajcr6/}.
The source code for reproducing the numerical results and figures in Sections \ref{sec:framework}, \ref{sec:discrete-stats}, and \ref{sec:case-study} is publicly available at 
\if1\anon {\url{https://github.com/eastonhuch/bayesian-randomization-inference}} \fi
\if0\anon {\url{https://github.com/redacted-for-peer-review}} \fi
.

\phantomsection\label{supplementary-material}
\bigskip

\begin{center}

{\large\bf SUPPLEMENTARY MATERIAL}

\end{center}

\begin{description}
\item[Online Supplement:]
Proofs, additional data analysis, and extensions. (PDF)
\end{description}

\bibliography{references}

\begin{thebibliography}{}

\bibitem[Bang and Robins, 2005]{bang_doubly_2005}
Bang, H. and Robins, J.~M. (2005).
\newblock Doubly robust estimation in missing data and causal inference models.
\newblock {\em Biometrics}, 61(4):962--973.

\bibitem[Berger and Boos, 1994]{berger_p_1994}
Berger, R.~L. and Boos, D.~D. (1994).
\newblock P values maximized over a confidence set for the nuisance parameter.
\newblock {\em Journal of the American Statistical Association}, 89(427):1012--1016.

\bibitem[Chen et~al., 2020]{chen_doubly_2020}
Chen, Y., Li, P., and Wu, C. (2020).
\newblock Doubly robust inference with nonprobability survey samples.
\newblock {\em Journal of the American Statistical Association}, 115(532):2011--2021.

\bibitem[Federer, 1969]{federer_geometric_1969}
Federer, H. (1969).
\newblock {\em Geometric Measure Theory}.
\newblock Classics in Mathematics. Springer Berlin Heidelberg, Berlin, Heidelberg.

\bibitem[Garthwaite, 1996]{garthwaite_confidence_1996}
Garthwaite, P.~H. (1996).
\newblock Confidence intervals from randomization tests.
\newblock {\em Biometrics}, 52(4):1387--1393.

\bibitem[Gelman and Shalizi, 2013]{gelman_philosophy_2013}
Gelman, A. and Shalizi, C.~R. (2013).
\newblock Philosophy and the practice of {B}ayesian statistics.
\newblock {\em British Journal of Mathematical and Statistical Psychology}, 66(1):8--38.

\bibitem[Gelman et~al., 2005]{gelman_multiple_2005}
Gelman, A., Van~Mechelen, I., Verbeke, G., Heitjan, D.~F., and Meulders, M. (2005).
\newblock Multiple imputation for model checking: Completed‐data plots with missing and latent data.
\newblock {\em Biometrics}, 61(1):74--85.

\bibitem[Greenland, 2005]{greenland_multiple-bias_2005}
Greenland, S. (2005).
\newblock Multiple-bias modelling for analysis of observational data.
\newblock {\em Journal of the Royal Statistical Society Series A: Statistics in Society}, 168(2):267--306.

\bibitem[Kennedy, 2023]{kennedy_towards_2023}
Kennedy, E.~H. (2023).
\newblock Towards optimal doubly robust estimation of heterogeneous causal effects.
\newblock {\em Electronic Journal of Statistics}, 17(2):3008--3049.

\bibitem[Luo et~al., 2021]{luo_leveraging_2021}
Luo, X., Dasgupta, T., Xie, M., and Liu, R.~Y. (2021).
\newblock Leveraging the {F}isher randomization test using confidence distributions: Inference, combination and fusion learning.
\newblock {\em Journal of the Royal Statistical Society Series B: Statistical Methodology}, 83(4):777--797.

\bibitem[McCandless and Gustafson, 2017]{mccandless_comparison_2017}
McCandless, L.~C. and Gustafson, P. (2017).
\newblock A comparison of {B}ayesian and {M}onte {C}arlo sensitivity analysis for unmeasured confounding.
\newblock {\em Statistics in Medicine}, 36(18):2887--2901.

\bibitem[Nie and Wager, 2021]{nie_quasi-oracle_2021}
Nie, X. and Wager, S. (2021).
\newblock Quasi-oracle estimation of heterogeneous treatment effects.
\newblock {\em Biometrika}, 108(2):299--319.

\bibitem[Robins et~al., 2000]{miller_sensitivity_2000}
Robins, J.~M., Rotnitzky, A., and Scharfstein, D.~O. (2000).
\newblock Sensitivity analysis for selection bias and unmeasured confounding in missing data and causal inference models.
\newblock In Miller, W., Halloran, M.~E., and Berry, D., editors, {\em Statistical Models in Epidemiology, the Environment, and Clinical Trials}, volume 116, pages 1--92. Springer New York, New York, NY.
\newblock Series Title: The IMA Volumes in Mathematics and its Applications.

\bibitem[Steenland, 2004]{steenland_monte_2004}
Steenland, K. (2004).
\newblock Monte {C}arlo sensitivity analysis and {B}ayesian analysis of smoking as an unmeasured confounder in a study of silica and lung cancer.
\newblock {\em American Journal of Epidemiology}, 160(4):384--392.

\bibitem[Wu and Thompson, 2020]{wu_sampling_2020}
Wu, C. and Thompson, M.~E. (2020).
\newblock {\em Sampling Theory and Practice}.
\newblock {ICSA} Book Series in Statistics. Springer International Publishing, Cham.

\end{thebibliography}


\begin{thebibliography}{}

\bibitem[Beaumont, 2019]{beaumont_approximate_2019}
Beaumont, M.~A. (2019).
\newblock Approximate {B}ayesian computation.
\newblock {\em Annual Review of Statistics and Its Application}, 6(1):379--403.

\bibitem[Bickel and Kleijn, 2012]{bickel_semiparametric_2012}
Bickel, P.~J. and Kleijn, B. J.~K. (2012).
\newblock The semiparametric {B}ernstein--von {M}ises theorem.
\newblock {\em The Annals of Statistics}, 40(1):206--237.

\bibitem[Bingham et~al., 2019]{bingham_pyro_2019}
Bingham, E., Chen, J.~P., Jankowiak, M., Obermeyer, F., Pradhan, N., Karaletsos, T., Singh, R., Szerlip, P., Horsfall, P., and Goodman, N.~D. (2019).
\newblock Pyro: Deep universal probabilistic programming.
\newblock {\em Journal of Machine Learning Research}, 20(1):973--978.

\bibitem[Bochkina and Green, 2014]{bochkina_bernsteinvon_2014}
Bochkina, N.~A. and Green, P.~J. (2014).
\newblock The {B}ernstein--von {M}ises theorem and nonregular models.
\newblock {\em The Annals of Statistics}, 42(5):1850--1878.

\bibitem[Chib and Hamilton, 2002]{CHIB200267}
Chib, S. and Hamilton, B.~H. (2002).
\newblock Semiparametric bayes analysis of longitudinal data treatment models.
\newblock {\em Journal of Econometrics}, 110(1):67--89.

\bibitem[Chiba, 2018]{chiba_bayesian_2018}
Chiba, Y. (2018).
\newblock Bayesian inference of causal effects for an ordinal outcome in randomized trials.
\newblock {\em Journal of Causal Inference}, 6(2):20170019.

\bibitem[Chung and Romano, 2013]{10.1214/13-AOS1090}
Chung, E. and Romano, J.~P. (2013).
\newblock {Exact and asymptotically robust permutation tests}.
\newblock {\em The Annals of Statistics}, 41(2):484 -- 507.

\bibitem[Ding, 2017]{ding_paradox_2017}
Ding, P. (2017).
\newblock A paradox from randomization-based causal inference.
\newblock {\em Statistical Science}, 32(3).

\bibitem[Ding and Guo, 2023]{DingGuo2023}
Ding, P. and Guo, T. (2023).
\newblock Posterior predictive propensity scores and p-values.
\newblock {\em Observational Studies}, 9(1):3--18.

\bibitem[Ding and Li, 2018]{ding_causal_2018}
Ding, P. and Li, F. (2018).
\newblock Causal inference: A missing data perspective.
\newblock {\em Statistical Science}, 33(2):214--237.

\bibitem[Ding and Miratrix, 2019]{ding_modelfree_2019}
Ding, P. and Miratrix, L.~W. (2019).
\newblock Model‐free causal inference of binary experimental data.
\newblock {\em Scandinavian Journal of Statistics}, 46(1):200--214.

\bibitem[Fisher, 1935]{fisher_design_1935}
Fisher, R.~A. (1935).
\newblock {\em The Design of Experiments}.
\newblock Oliver \& Boyd.

\bibitem[Gelman et~al., 2014]{gelman_bayesian_2014}
Gelman, A., Carlin, J.~B., Stern, H.~S., Dunson, D.~B., Vehtari, A., and Rubin, D.~B. (2014).
\newblock {\em Bayesian Data Analysis}.
\newblock Texts in Statistical Science Series. CRC Press, Taylor and Francis Group, Boca Raton London New York, third edition.

\bibitem[Gelman et~al., 1996]{gelman_posterior_1996}
Gelman, A., Meng, X.-L., and Stern, H. (1996).
\newblock Posterior predictive assessment of model fitness via realized discrepancies.
\newblock {\em Statistica Sinica}, 6:733--760.

\bibitem[Gelman and Rubin, 1992]{gelman_inference_1992}
Gelman, A. and Rubin, D.~B. (1992).
\newblock Inference from iterative simulation using multiple sequences.
\newblock {\em Statistical Science}, 7(4):457--472.

\bibitem[Ghosh and Ramamoorthi, 2003]{ghosh_bayesian_2003}
Ghosh, J.~K. and Ramamoorthi, R.~V. (2003).
\newblock {\em Bayesian Nonparametrics}.
\newblock Springer Series in Statistics. Springer-Verlag, New York.

\bibitem[Gutmann and Corander, 2016]{gutmann_bayesian_2016}
Gutmann, M.~U. and Corander, J. (2016).
\newblock Bayesian optimization for likelihood-free inference of simulator-based statistical models.
\newblock {\em Journal of Machine Learning Research}, 17(125):1--47.

\bibitem[Hodges and Lehmann, 1963]{hodges_estimates_1963}
Hodges, J.~L. and Lehmann, E.~L. (1963).
\newblock Estimates of location based on rank tests.
\newblock {\em The Annals of Mathematical Statistics}, 34(2):598--611.

\bibitem[Holland, 1986]{holland_statistics_1986}
Holland, P.~W. (1986).
\newblock Statistics and causal inference (with discussion).
\newblock {\em Journal of the American Statistical Association}, 81(396):945--960.

\bibitem[Humphreys and Jacobs, 2015]{humphreys_mixing_2015}
Humphreys, M. and Jacobs, A.~M. (2015).
\newblock Mixing methods: A {B}ayesian approach.
\newblock {\em American Political Science Review}, 109(4):653--673.

\bibitem[Imbens and Rubin, 2015]{Imbens_Rubin_2015}
Imbens, G.~W. and Rubin, D.~B. (2015).
\newblock {\em Causal Inference for Statistics, Social, and Biomedical Sciences: An Introduction}.
\newblock Cambridge University Press.

\bibitem[Keele and Quinn, 2017]{keele_bayesian_2017}
Keele, L. and Quinn, K.~M. (2017).
\newblock Bayesian sensitivity analysis for causal effects from $2 \times 2$ tables in the presence of unmeasured confounding with application to presidential campaign visits.
\newblock {\em The Annals of Applied Statistics}, 11(4).

\bibitem[Kim, 2002]{kim_limited_2002}
Kim, J.-Y. (2002).
\newblock Limited information likelihood and {B}ayesian analysis.
\newblock {\em Journal of Econometrics}, 107(1):175--193.

\bibitem[Kleijn and van~der Vaart, 2012]{kleijn_bernstein-von-mises_2012}
Kleijn, B. and van~der Vaart, A. (2012).
\newblock The {B}ernstein--{V}on--{M}ises theorem under misspecification.
\newblock {\em Electronic Journal of Statistics}, 6:354--381.

\bibitem[Kwan, 1999]{kwan_asymptotic_1999}
Kwan, Y.~K. (1999).
\newblock Asymptotic {B}ayesian analysis based on a limited information estimator.
\newblock {\em Journal of Econometrics}, 88(1):99--121.

\bibitem[Leavitt, 2023]{leavitt_randomization-based_2023}
Leavitt, T. (2023).
\newblock Randomization-based, {B}ayesian inference of causal effects.
\newblock {\em Journal of Causal Inference}, 11(1):20220025.

\bibitem[Li et~al., 2023]{li_bayesian_2023}
Li, F., Ding, P., and Mealli, F. (2023).
\newblock Bayesian causal inference: A critical review.
\newblock {\em Philosophical Transactions of the Royal Society A: Mathematical, Physical and Engineering Sciences}, 381(2247):20220153.

\bibitem[Li et~al., 2018]{li_balancing_2018}
Li, F., Morgan, K.~L., and Zaslavsky, A.~M. (2018).
\newblock Balancing covariates via propensity score weighting.
\newblock {\em Journal of the American Statistical Association}, 113(521):390--400.

\bibitem[Li and Ding, 2017]{li_general_2017}
Li, X. and Ding, P. (2017).
\newblock General forms of finite population central limit theorems with applications to causal inference.
\newblock {\em Journal of the American Statistical Association}, 112(520):1759--1769.

\bibitem[Marty et~al., 2020]{marty_socioeconomic_2020}
Marty, L., Jones, A., and Robinson, E. (2020).
\newblock Socioeconomic position and the impact of increasing availability of lower energy meals vs. menu energy labelling on food choice: Two randomized controlled trials in a virtual fast-food restaurant.
\newblock {\em International Journal of Behavioral Nutrition and Physical Activity}, 17(1):1--11.

\bibitem[Meng, 1994]{meng_posterior_1994}
Meng, X.-L. (1994).
\newblock Posterior predictive $p$-values.
\newblock {\em The Annals of Statistics}, 22(3):1142--1160.

\bibitem[Miller, 2021]{miller_asymptotic_2021}
Miller, J.~W. (2021).
\newblock Asymptotic normality, concentration, and coverage of generalized posteriors.
\newblock {\em Journal of Machine Learning Research}, 22(168):1--53.

\bibitem[Miller and Dunson, 2019]{miller_robust_2019}
Miller, J.~W. and Dunson, D.~B. (2019).
\newblock Robust {B}ayesian inference via coarsening.
\newblock {\em Journal of the American Statistical Association}, 114(527):1113--1125.

\bibitem[Neyman, 1990]{splawa-neyman_application_1990}
Neyman, J. (1923/1990).
\newblock On the application of probability theory to agricultural experiments. essay on principles. section 9.
\newblock {\em Statistical Science}, 5(4):465--472.
\newblock Translated from the 1923 Polish original and edited by D. M. Dabrowska and T. P. Speed.

\bibitem[Robins and Ritov, 1997]{robins_toward_1997}
Robins, J.~M. and Ritov, Y. (1997).
\newblock Toward a curse of dimensionality appropriate ({CODA}) asymptotic theory for semi-parametric models.
\newblock {\em Statistics in Medicine}, 16(3):285--319.

\bibitem[Romano, 1990]{Romano01091990}
Romano, J.~P. (1990).
\newblock On the behavior of randomization tests without a group invariance assumption.
\newblock {\em Journal of the American Statistical Association}, 85(411):686--692.

\bibitem[Rosenbaum, 1993]{rosenbaum_hodges-lehmann_1993}
Rosenbaum, P.~R. (1993).
\newblock Hodges--{L}ehmann point estimates of treatment effect in observational studies.
\newblock {\em Journal of the American Statistical Association}, 88(424):1250--1253.

\bibitem[Rosenbaum, 2002]{rosenbaum_observational_2002}
Rosenbaum, P.~R. (2002).
\newblock {\em Observational Studies}.
\newblock Springer Series in Statistics. Springer, New York, NY, second edition.

\bibitem[Rousseau, 2016]{rousseau_frequentist_2016}
Rousseau, J. (2016).
\newblock On the frequentist properties of {B}ayesian nonparametric methods.
\newblock {\em Annual Review of Statistics and Its Application}, 3(1):211--231.

\bibitem[Rubin, 1978]{rubin_bayesian_1978}
Rubin, D.~B. (1978).
\newblock Bayesian inference for causal effects: The role of randomization.
\newblock {\em The Annals of Statistics}, 6(1):34--58.

\bibitem[Rubin, 1985]{Rubin1985Propensity}
Rubin, D.~B. (1985).
\newblock The use of propensity score in applied {B}ayesian inference.
\newblock In Bernardo, J.~M., DeGroot, M.~H., Lindley, D.~V., and Smith, A. F.~M., editors, {\em Bayesian Statistics, Volume 2}, pages 463--472. North-Holland: Elsevier Science Publishers B.V., Amsterdam.

\bibitem[Saarela et~al., 2016]{saarela_bayesian_2016}
Saarela, O., Belzile, L.~R., and Stephens, D.~A. (2016).
\newblock A {Bayesian} view of doubly robust causal inference.
\newblock {\em Biometrika}, 103(3):667--681.

\bibitem[Shi and Ding, 2023]{shi_Berry--Esseen_2023}
Shi, L. and Ding, P. (2023).
\newblock {B}erry--{E}sseen bounds for design-based causal inference with possibly diverging treatment levels and varying group sizes.
\newblock arXiv:2209.12345.

\bibitem[van~der Vaart, 1998]{vaart_asymptotic_1998}
van~der Vaart, A.~W. (1998).
\newblock {\em Asymptotic Statistics}.
\newblock Cambridge University Press, first edition.

\bibitem[Wilcoxon, 1945]{c4091bd3-d888-3152-8886-c284bf66a93a}
Wilcoxon, F. (1945).
\newblock Individual comparisons by ranking methods.
\newblock {\em Biometrics Bulletin}, 1(6):80--83.

\bibitem[Wood, 2010]{WoodSimonN.2010Sifn}
Wood, S.~N. (2010).
\newblock Statistical inference for noisy nonlinear ecological dynamic systems.
\newblock {\em Nature (London)}, 466(7310):1102--1104.

\bibitem[Zellner, 1986]{zellner_assessing_1986}
Zellner, A. (1986).
\newblock On assessing prior distributions and {B}ayesian regression analysis with g prior distributions.
\newblock In Goel, P. and Zellner, A., editors, {\em Bayesian Inference and Decision Techniques: Essays in Honor of {B}runo de {F}inetti}, volume~6 of {\em Studies in {B}ayesian Econometrics and Statistics}, pages 233--243. Elsevier, New York.

\end{thebibliography}

\clearpage
\appendix
\setcounter{page}{1}
\pagenumbering{arabic} 

\spacingset{1}

\if1\anon
{
  \begin{center}
  \LARGE \bf Supplementary Material for \\[1ex]
  ``Robust Bayesian Inference of Causal Effects via Randomization Distributions'' \\[3ex]
  \normalfont \large
  Easton Huch\\
  Department of Statistics, University of Michigan\\[2ex]
  Fred Feinberg\\
  Department of Marketing, University of Michigan\\[2ex]
  Walter Dempsey\\
  Department of Biostatistics, University of Michigan\\[4ex]
  \today
  \end{center}
} \fi

\if0\anon
{
  \bigskip
  \bigskip
  \bigskip
  \begin{center}
    {\LARGE \bf Supplementary Material for \\[1ex]
  ``Robust Bayesian Inference of Causal Effects via Randomization Distributions'' \\[4ex]
  \normalfont \large
  \today}
  \end{center}
  \medskip
} \fi

\vspace{3ex}

\section{Simulation Details}\label{app:simulation-details}

This section provides additional results from the simulations of Section \ref{sec:discrete-simulation} and details on the simulation setup.
The simulation repetitions consist of the following steps:
\begin{enumerate}
    \item Set $y_{0i} = z_i + g_i$, where $z_i \overset{iid}{\sim} \Normaldist(0, 10^2)$ and $g_i \overset{iid}{\sim} \Gammadist(4, 2.5)$.
    \item Sample $\theta \sim \Normaldist(0, 10^2)$.
    \item Set $y_{1i} = y_{0i} + \theta$.
    \item Sample $\vec{a}$ according to complete randomization with $\sum_{i=1}^n a_i \eqqcolon n_1 = n_0 \coloneqq \sum_{i=1}^n (1 - a_i)$; i.e., all values of $\vec{a}$ resulting in $n_0 = n_1$ are equally likely.
    \item Set $\vec{y}_{\vec{a}} = \vec{a} \odot \vec{y}_1 + (\vec{1}_n - \vec{a}) \odot \vec{y}_0$, where $\odot$ denotes elementwise multiplication.
    \item Compute the statistic, $s$.
    \item Compute the posterior mean of $\theta$ and a centered 95\% CI.
\end{enumerate}
We repeat this process for the six methods described in Section \ref{sec:discrete-simulation}.
In this appendix, we include two additional methods:
\begin{itemize}
    \item BRI-R: BRI model with a DIM statistic \textbf{R}ounded to the nearest integer.
    \item BRI-U: \textbf{U}nidirectional BRI model $y_{1i} = y_{0i} + \theta + \epsilon_i$, $\epsilon_i \overset{iid}{\sim} \Normaldist(0, 1)$ and statistic $s_1$.
\end{itemize}
We also test oracle versions of the BRI methods in which the model observes a statistic generated from a new, independent assignment vector that (potentially) differs from that corresponding to $\vec{y_a}$.
These oracle methods, while not computable in practice, offer an interesting comparison because they allow us to assess how reusing $\vec{a}$ affects the performance of the BRI methods.
We denote the oracle methods with an asterisk (*) after their name, such as BRI-U*.

Table \ref{tab:discrete-exact-app} shows the results of Table \ref{tab:discrete-exact} with these additional methods.
The estimates and inferences produced by BRI-R and BRI-U perform similarly to those from the BRI-C and BRI-A methods.
The performance of the oracle methods is similar to that of the standard BRI methods, except they produce coverage rates within Monte Carlo error of the nominal 95\% level.
Traditional Bayesian methods should have exact coverage guarantees with the parameter drawn from the prior, highlighting how BRI falls short of being fully Bayesian.
Nonetheless, BRI can be regarded as an approximation to these oracle methods, and the theoretical results in Section \ref{sec:theory} demonstrate that, in regular parametric settings, this is purely a finite-sample phenomenon.

\begin{table}[htb]
\centering
\begin{tabular}{lrrrr}
\toprule
\textbf{Metrics} & \multicolumn{1}{c}{\textbf{Bias}} & \multicolumn{1}{c}{\textbf{MSE}} & \multicolumn{1}{c}{\textbf{Coverage}} & \multicolumn{1}{c}{\textbf{CI Length}} \\
\midrule
Prior & 0.060 (0.100) & 100.802 (1.431) & 0.950 (0.002) & 39.210 (0.000) \\
DIM & -0.063 (0.071) & 49.968 (0.710) & 0.882 (0.003) & 24.104 (0.062) \\
LIB & -0.031 (0.059) & 35.338 (0.504) & 0.893 (0.003) & 20.155 (0.038) \\
BRI-U & -0.004 (0.066) & 43.293 (0.736) & 0.965 (0.002) & 28.635 (0.051) \\
BRI-U* & 0.044 (0.063) & 39.626 (0.619) & 0.951 (0.002) & 25.144 (0.044) \\
BRI-A & -0.013 (0.062) & 38.725 (0.613) & 0.959 (0.002) & 26.763 (0.046) \\
BRI-A* & 0.036 (0.063) & 39.669 (0.636) & 0.955 (0.002) & 25.706 (0.043) \\
BRI-R & 0.002 (0.070) & 48.550 (0.848) & 0.967 (0.002) & 30.386 (0.051) \\
BRI-R* & 0.036 (0.063) & 39.309 (0.616) & 0.949 (0.002) & 24.930 (0.045) \\
BRI-C & 0.002 (0.070) & 48.488 (0.847) & 0.967 (0.002) & 30.385 (0.051) \\
BRI-C* & 0.043 (0.063) & 39.453 (0.621) & 0.950 (0.002) & 24.948 (0.045) \\
BRI-RS & -0.028 (0.063) & 39.589 (0.580) & 0.980 (0.001) & 30.668 (0.043) \\
BRI-RS* & 0.048 (0.066) & 43.080 (0.624) & 0.951 (0.002) & 26.893 (0.046) \\
\bottomrule
\end{tabular}
\caption{\label{tab:discrete-exact-app} Empirical bias, MSE, 95\% CI coverage, and average 95\% CI length for the methods in the first simulation study. The values in parentheses denote estimated Monte Carlo standard errors. Compare to Table \ref{tab:discrete-exact}.}
\end{table}

\begin{figure}[htbp]
\centering
\begin{subfigure}{0.32\textwidth}
\includegraphics[width=\linewidth]{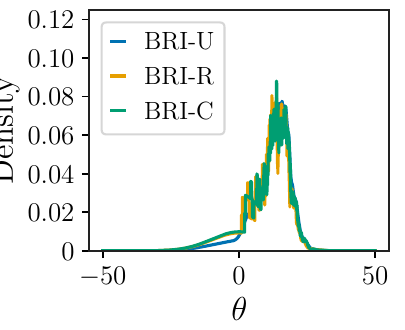}
\caption{Discretization Strategies}
\label{fig:discrete-stats-exact:dim}
\end{subfigure}
\begin{subfigure}{0.32\textwidth}
\includegraphics[width=\linewidth]{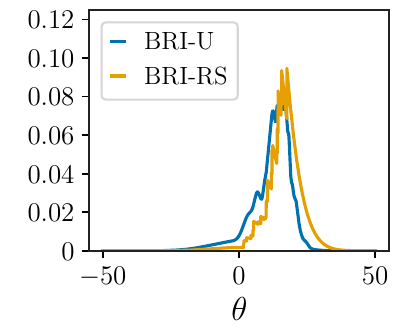}
\caption{RS Statistic}
\label{fig:discrete-stats-exact:rs}
\end{subfigure}
\begin{subfigure}{0.32\textwidth}
\includegraphics[width=\linewidth]{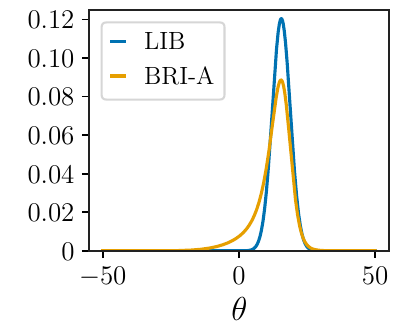}
\caption{Asymptotic Approximations}
\label{fig:discrete-stats-exact:asymp}
\end{subfigure}
\caption{Comparison of posterior distributions for selected methods from the discrete statistic simulation study. Panel (a) compares three strategies for handling the discreteness of the DIM statistic, resulting in similar inferences. Panel (b) compares one of the methods in Panel (a) to a BRI model with an RS statistic (BRI-RS); the inferences differ relatively more compared to Panel (a) because these statistics contain different information. Panel (c) compares an LIB approach to an asymptotic approximation to the BRI likelihood (BRI-A); whereas the LIB posterior is constrained to be symmetric, the BRI-A posterior is not, potentially explaining its superior performance in the simulation study.}
\label{fig:discrete-stats-exact}
\end{figure}

Figure \ref{fig:discrete-stats-exact} plots posterior distributions from a single repetition of the first simulation study.
Panel (a) compares BRI-R, BRI-C, and BRI-U, all of which result in similar posterior distributions; in particular, BRI-C and BRI-U are visually indistinguishable.
Panel (b) compares BRI-U and BRI-RS, highlighting the differences between the inferences produced by the RS statistic and those based on sample means.
Panel (c) compares BRI-A and LIB.
Whereas LIB is constrained to a symmetric Gaussian posterior, BRI-A produces a data-adapted asymmetric posterior distribution.

Due to computational constraints, we did not include the BRI-R and BRI-U methods in the second simulation.
One additional detail regarding this simulation is that the nominal coverage levels are not exactly equal to 0.950 due to the discretization of the parameter space;
however, they always fall in the range $(0.950, 0.952)$, so this detail has a negligible impact on the results shown in Table \ref{tab:discrete-simulation-asymp}.

\section{Model Checking in BRI}\label{app:model-checking}

This appendix provides additional details on model checking  within the BRI framework.
In Bayesian philosophy and practice, model checking is increasingly viewed as an integral part of the scientific process that enables exploration and adoption of models with increasing explanatory power \citepApp{gelman_philosophy_2013}.
Within the BRI framework, the model-checking process is facilitated by the fact that the likelihood function is precisely the same distribution that would be used in an FRT for the same model and statistic.
Both Bayesian model checking and FRTs often employ discrepancy variables---a connection we highlight below.

\subsection{Discrepancy Variables}
\label{sec:discrepancy-variables}

A \emph{discrepancy variable} (or simply \emph{discrepancy}) generalizes the definition of a statistic to allow dependence on parameters in addition to data \citep{gelman_posterior_1996}.
This generalization is natural in the Bayesian paradigm because both data and parameters are viewed as random variables.
The following example illustrates how a discrepancy variable could be used to check a modeling assumption in the standard superpopulation framework.

\begin{example}
Suppose $z_i \overset{iid}{\sim} \text{Normal}(\mu, 1)$.
Because the Gaussian distribution is symmetric, we may desire to check deviations from the model in terms of skew.
A natural discrepancy variable for this objective is
\[d_{\mu} =\left|\frac{1}{n} \sum_{i=1}^n (z_i - \mu)^3\right|\]
with larger values indicating greater evidence against the Gaussian assumption.
\end{example}

In practice, we can compute $d_{\mu}$ for a set of posterior samples of $\mu$.
Comparing these samples to simulated values from the posterior predictive distribution produces a measure of extremeness of the observed data relative to the assumed model---a ``posterior predictive $p$-value'' \citep{meng_posterior_1994}.
Small posterior predictive $p$-values provide evidence against the assumed model and often suggest directions in which to generalize it, such as replacing the Gaussian distribution with a skewed distribution.

\subsection{Embedded FRTs}
\label{sec:frts}

The logic of FRTs is similar to that of posterior predictive $p$-values in that both rely on measures of extremeness to quantify evidence against an assumed model.
In addition, FRTs may also rely on discrepancy variables, as the following example illustrates.

\begin{example}
\label{ex:frt-diff-means}
Assume $\vec{a}$ is completely randomized so that all values of $\vec{a}$ having $\sum_{i=1}^n a_i \eqqcolon n_1 \in \N$ are equally likely,
and suppose that interest lies in the constant treatment effect model: $y_{1i} = y_{0i} + \theta$.
For any given value of $\theta$, we can form a hypothesis test by
\begin{enumerate}[label=(\alph*)]
    \item imputing the counterfactuals,
    \item calculating the randomization distribution for a prespecified statistic, and
    \item comparing the observed value of the same statistic to its randomization distribution, resulting in a $p$-value.
\end{enumerate}
In the logic of hypothesis testing, a simple hypothesis $\text{H}_{\theta}$ fixes the value of $\theta$.
Thus, under $\text{H}_{\theta}$, we would be justified in replacing the statistic with a discrepancy variable that relies on imputed counterfactuals and/or $\theta$.
For example, instead of the DIM statistic $s_{\Delta}$, we could conduct an FRT in terms of a difference-in-control-means discrepancy variable $d_{\Delta0} \coloneqq d_0 - s_0$, where
\[
d_0 \coloneqq \frac{\sum_{i=1}^n a_i (y_{ai} - a_i\theta)}{\sum_{i=1}^n 
a_i} - s_0,
\]
and $y_{ai} - a_i\theta$ is a model-based imputation of $y_{0i}$.
In fact, these two formulations are easily seen to produce identical $p$-values.
\end{example}

In addition to testing prespecified hypotheses, we may also invert a sequence of FRTs to form a confidence interval for $\theta$ \citepApp{garthwaite_confidence_1996, luo_leveraging_2021}.
In Example \ref{ex:frt-diff-means}, setting $\theta = s_{\Delta}$ will result in no evidence against the model.
Thus, these discrepancy variables can distinguish between values of $\theta$, but they do not provide evidence to falsify the constant treatment effect model, which may be a separate objective.
The example below shows how to accomplish this objective in a continuation of Example \ref{ex:frt-diff-means}.

\begin{example}[Continuation of Example \ref{ex:frt-diff-means}]
\label{ex:diff-in-variance}
Define the statistic $s_2 \coloneqq |\log(s_{12}/s_{02})|$, where $s_{12}$ is defined in \eqref{eq:s12} and
\[
    s_{02} \coloneq
    \frac{\sum_{i=1}^n (1 - a_i) (y_{ai} - s_0)^2}{\sum_{i=1}^n 
    (1 - a_i)},
\]
so that $s_{12}/s_{02}$ is the ratio of sample variances between the treatment and control groups.
Under the constant treatment effect model, we expect $s_2$ to be close to zero because the model implies that the treatment and control groups have equal variances.
\end{example}

Within the FRT framework, we could perform a sequence of tests over all $\theta$ and compute the maximum $p$-value, or apply the method of \citetApp{berger_p_1994}.
Alternatively, we could compute two $1-\alpha/2$ confidence intervals using $s_{\Delta}$ and $s_2$, respectively, and construct a $1 - \alpha$ confidence interval from their intersection.
A small $p$-value or empty interval would indicate evidence against the constant treatment effect model, leading us to alternative theories regarding the form of the causal effects.

In a similar fashion, we can compute posterior predictive $p$-values using $s_{\Delta}$, $s_2$, or any other discrepancy variable within the BRI framework.
In cases where we approximate the likelihood function via Monte Carlo simulation, we can simply reuse the sampled values of $\vec{a}$ to compute the posterior predictive $p$-value.
Under stochastic treatment effect models, this process can be further augmented by sampling counterfactual outcomes \citepApp{gelman_multiple_2005}.
Because this model-checking process applies the FRT using posterior samples for $\vec{\theta}$, it results in a posterior predictive distribution that averages over uncertainty in nuisance variables.
Our case study in Section \ref{sec:case-study} provides an example of this model-checking process.

\subsection{Inference via Discrepancy Variables}
\label{sec:inference-discrepancies}

Because FRTs can be conducted directly using discrepancy variables, we may ask the question: Can we apply BRI directly to discrepancy variables?
To facilitate the discussion, we consider the case where the discrepancy variable can be represented as a bi-Lipschitz map, $\vec{d}_{\vec{\theta}}: \R^k \to \R^k$, of the statistic, $\vec{s}$.
We further assume that $\vec{s}$ follows a distribution that is absolutely continuous with respect to Lebesgue measure.
By the change-of-variables formula, we can relate the densities as
\begin{equation}
\label{eq:disc-density}
p(\vec{s} | \vec{\theta}, \vec{y_a}) = p\{\vec{d}_{\vec{\theta}}(\vec{s}) | \vec{\theta}, \vec{y_a}\} \cdot \left|\det \vec{d}'_{\vec{\theta}}(\vec{s})\right|,
\end{equation}
where $\vec{d}'_{\vec{\theta}}(\vec{s})$ is the (Jacobian) derivative of $\vec{d}_{\vec{\theta}}(\vec{s})$.
So, supposing we calculate $p\{\vec{d}_{\vec{\theta}}(\vec{s}) | \vec{\theta}, \vec{y_a}\}$, we can recover the standard analysis by multiplying this discrepancy density by the Jacobian factor on the right-hand side of \eqref{eq:disc-density}.
For $s_{\Delta}$ and $d_{\Delta0}$, we have $d_{\Delta0} = s_{\Delta} - \theta$ so that the Jacobian factor is unity and $p(\vec{s} | \vec{\theta}, \vec{y_a}) = p\{\vec{d}_{\vec{\theta}}(\vec{s}) | \vec{\theta}, \vec{y_a}\}$.

For more general discrepancy variables, the relationship between the corresponding integrals is more complicated, and there may not exist a simple transformation of the discrepancy density to the statistic density; instead, they are related by the area and coarea formulas in \citetApp{federer_geometric_1969}.
However, provided we can express the discrepancy variable in terms of a statistic, we can directly compute the likelihood using the statistic itself without the need to first calculate $p\{\vec{d}_{\vec{\theta}}(\vec{s}) | \vec{\theta}, \vec{y_a}\}$.

\section{Proofs of Theoretical Results}\label{app:proofs}

This appendix provides proofs of several theoretical results in Section \ref{sec:theory}.

\subsection{Proof of Lemma \ref{lemma:likelihood}}
\label{sec:proof-likelihood}

The observed event is $\{\vec{s}_n \in \mathcal{N}_{\epsilon_n}(\vec{s}_n^*)\} = \{\|\vec{s}_n - \vec{s}_n^*\|_{\infty} \leq \epsilon_n\}$.
Multiplying by $\sqrt{n}$ inside the probability statement and centering by $\vec{r}_n(\vec{\theta})$, we obtain
\[
\Pr\left(\left\|\sqrt{n} \left[\vec{s}_n - \vec{r}_n(\vec{\theta}) - \left\{\vec{s}_n^* - \vec{r}_n(\vec{\theta})\right\}\right]\right\|_{\infty} \leq \epsilon_n \sqrt{n} | \vec{\theta}, \vec{y}_{\vec{a}n}\right).
\]
Letting $\vec{z}_{\vec{\theta}n} \sim \Normaldist\{\vec{0}, \mat{V}_n(\vec{\theta})\}$ and applying the Gaussian approximation of Assumption \ref{assumption:clt} gives
\begin{equation}
\label{eq:likelihood-bound}
\left|\Pr\left\{\vec{s}_n \in \mathcal{N}_{\epsilon_n}(\vec{s}_n^*)| \vec{\theta}, \vec{y}_{\vec{a}n}\right\} - \Pr\left[\left\| \vec{z}_{\vec{\theta}n} - \sqrt{n} \left\{\vec{s}_n^* - \vec{r}_n(\vec{\theta})\right\}\right\|_{\infty} \leq \epsilon_n \sqrt{n} | \vec{\theta}, \vec{y}_{\vec{a}n}\right] \right| 
\leq C_1n^{-1/2},
\end{equation}
for some $C_1 > 0$ with probability at least $1 - \delta/2$ for $n \geq N_1 \in \N$.
To obtain a nonzero limit, we normalize the approximation, dividing by $(2\epsilon_n \sqrt{n})^p$.
The bound on the right then becomes
\[
\frac{C_1 n^{-1/2}}{(2\epsilon_n \sqrt{n})^p}
= 2^{-p} C_1 n^{p(\alpha - 1/2) -1/2},
\]
which is $o(1)$ for $\alpha < (p+1) / (2p)$.
We then note that the probability involving $\vec{z}_{\vec{\theta}n}$ converges to the Gaussian probability density.
To see this, let $\vec{b}_n \coloneqq \sqrt{n} \left\{\vec{s}_n^* - \vec{r}_n(\vec{\theta})\right\}$ and rewrite this term as an integral:
\begin{align*}
&\ (2\epsilon_n \sqrt{n})^{-p} \Pr\left(\left\| \vec{z}_{\vec{\theta}n} - \vec{b}_n\right\|_{\infty} \leq \epsilon_n \sqrt{n} | \vec{\theta}, \vec{y}_{\vec{a}n}\right) \\
=&\ (2\epsilon_n \sqrt{n})^{-p} \int_{b_{n1} - \epsilon_n \sqrt{n}}^{b_{n1} + \epsilon_n \sqrt{n}} \ldots \int_{b_{np} - \epsilon_n \sqrt{n}}^{b_{np} + \epsilon_n \sqrt{n}} \phi\left\{\vec{t}, \vec{0}, \mat{V}_n(\vec{\theta})\right\} d\vec{t} \\
=&\ (2\epsilon_n \sqrt{n})^{-p} \phi\left\{\vec{c}_n, \vec{0}, \mat{V}_n(\vec{\theta})\right\} \int_{b_{n1} - \epsilon_n \sqrt{n}}^{b_{n1} + \epsilon_n \sqrt{n}} \ldots \int_{b_{np} - \epsilon_n \sqrt{n}}^{b_{np} + \epsilon_n \sqrt{n}} d\vec{t} \\
=&\ \phi\left\{\vec{c}_n, \vec{0}, \mat{V}_n(\vec{\theta})\right\}
\end{align*}
for some $\vec{c}_n \in \{\vec{c} \in \R^p: \|\vec{b}_n - \vec{c}\|_{\infty} < \epsilon_n \sqrt{n}\}$ by the mean-value theorem for iterated integrals.
Then, applying the mean-value theorem for a function of multiple variables, we obtain
\[
\phi\left\{\vec{c}_n, \vec{0}, \mat{V}_n(\vec{\theta})\right\}
=
\phi\left\{\vec{b}_n, \vec{0}, \mat{V}_n(\vec{\theta})\right\} +
\vec{\phi}'\{(1-t) \vec{b}_n + t \vec{c}_n\} (\vec{c}_n - \vec{b}_n)
\]
for some $t \in (0, 1)$.
Then, by Cauchy--Schwarz, we have 
\[
|\vec{\phi}'\{(1-t) \vec{b}_n + t \vec{c}_n\} (\vec{c}_n - \vec{b}_n)|
\leq
\|\vec{\phi}'\{(1-t) \vec{b}_n + t \vec{c}_n\}\|_2 \cdot
\|\vec{c}_n - \vec{b}_n\|_2 \leq C_2 \epsilon_n \sqrt{n}
\]
for some $C_2 > 0$ with probability at least $1-\delta/2$ for $n \geq N_2 \in \N$ because Assumption \ref{assumption:V} implies that $\vec{\phi}'$ can be bounded with high probability for sufficiently large $n$.
Applying this bound to \eqref{eq:likelihood-bound} with the reverse triangle inequality yields
\[
\left|
\frac{\Pr\left\{\vec{s}_n \in \mathcal{N}_{\epsilon_n}(\vec{s}_n^*)| \vec{\theta}, \vec{y}_{\vec{a}n}\right\}}{(2\epsilon_n \sqrt{n})^p} -
\phi\{\vec{b}_n, \vec{0}, \mat{V}_n(\vec{\theta})\}
\right| 
\leq 2^{-p} C_1 n^{p(\alpha - 1/2) - 1/2} + C_2 n^{1/2 - \alpha}
\]
with probability at least $1 - \delta$ for $n \geq N \coloneqq \max(N_1, N_2)$.
Letting $C \coloneqq 2\max(C_1, C_2)$, the above bound is no greater than $Cn^{\gamma}$ as claimed in the lemma.
The bound is minimized by equating $p(\alpha - 1/2) - 1/2 = 1/2 - \alpha$, which gives
\[
\alpha = \frac{2+p}{2(p+1)},\quad \gamma = -\frac{1}{2(p+1)}.
\]
Although Lemma \ref{lemma:likelihood} employs the supremum norm to define neighborhoods, we could obtain similar results for other norms by bounding them via the supremum norm.


\subsection{Proof of Theorem \ref{thm:bvm}}
\label{sec:proof-bvm}

We provide a proof sketch.
The technical details may be adapted from \citet{vaart_asymptotic_1998}, \citet{ghosh_bayesian_2003}, and the references therein.
By Bayes' rule, the posterior density is
\[
p\left\{\vec{\theta} | \vec{y}_{\vec{a}n}, \vec{s}_n \in \mathcal{N}_{\epsilon_n}\left(\vec{s}_n^*\right)\right\}
= \frac{p(\vec{\theta} | \vec{y}_{\vec{a}n}) p\left\{\vec{s}_n \in \mathcal{N}_{\epsilon_n}\left(\vec{s}_n^*\right) | \vec{\theta}, \vec{y}_{\vec{a}n} \right\}}{k(\vec{s}^*_n, \vec{y}_{\vec{a}n})},
\]
where the normalizing constant, $k(\vec{s}^*_n, \vec{y}_{\vec{a}n})$, is defined as
\[
k(\vec{s}^*_n, \vec{y}_{\vec{a}n})
\coloneqq
\int_{\vec{\theta} \in \mathcal{T}}
p(\vec{\theta} | \vec{y}_{\vec{a}n}) p\left\{\vec{s}_n \in \mathcal{N}_{\epsilon_n}\left(\vec{s}_n^*\right) | \vec{\theta}, \vec{y}_{\vec{a}n} \right\} d\vec{\theta}.
\]
To obtain nonzero limits, we multiply and divide by $(2\epsilon_n \sqrt{n})^p$ as follows:
\[
p\left\{\vec{\theta} | \vec{y}_{\vec{a}n}, \vec{s}_n \in \mathcal{N}_{\epsilon_n}\left(\vec{s}_n^*\right)\right\} =
p(\vec{\theta} | \vec{y}_{\vec{a}n}) \cdot
\frac{p\left\{\vec{s}_n \in \mathcal{N}_{\epsilon_n}\left(\vec{s}_n^*\right) | \vec{\theta}, \vec{y}_{\vec{a}n} \right\}}{(2\epsilon_n \sqrt{n})^p} \cdot
\frac{(2\epsilon_n \sqrt{n})^p}{k(\vec{s}^*_n, \vec{y}_{\vec{a}n})}.
\]
We can then apply Lemma \ref{lemma:likelihood} to approximate $k(\vec{s}^*_n, \vec{y}_{\vec{a}n}) / (2\epsilon_n \sqrt{n})^p$.
In a sufficiently small neighborhood of $\vec{\theta}^*$, we can approximate $\mat{V}_n(\vec{\theta}) \approx \mat{V}(\vec{\theta}^*)$ by Assumption \ref{assumption:V} and $\vec{r}_n(\vec{\theta}) \approx \vec{r}(\vec{\theta}^*) + \vec{r}'(\vec{\theta}^*)(\vec{\theta} - \vec{\theta}^*)$, the latter following from Assumption \ref{assumption:r} and a Taylor-series expansion.
In this neighborhood, the Gaussian approximation from Lemma \ref{lemma:likelihood} is then $(2\pi)^{-p/2} \det\{\mat{V}(\vec{\theta}^*)\}^{-1/2} \exp\left\{-\left(
\vec{\theta} - \vec{\mu}_n
\right) \mat{\Sigma}^{-1} \left(
\vec{\theta} - \vec{\mu}_n
\right)
\cdot n/2\right\}$.
Outside this neighborhood, the likelihood converges to zero at an exponential rate.
Integrating over $\vec{\theta}$, we then have \[k(\vec{s}^*_n, \vec{y}_{\vec{a}n}) / (2\epsilon_n \sqrt{n})^p \overset{p}{\to} \sqrt{\det(\mat{\Sigma}) / \det\{\mat{V}(\vec{\theta}^*)\}} \cdot \plim p(\vec{\theta^* | \vec{y}_{\vec{a}n}}) \eqqcolon K\]
due to the compactness of $\mathcal{T}$ (Assumption \ref{assumption:compact}).
The term $\plim p(\vec{\theta^* | \vec{y}_{\vec{a}n}})$ is the limit of the prior density near $\vec{\theta}^*$ (Assumption \ref{assumption:prior}).
The contribution of the prior outside of this neighborhood is negligible due to (a) the exponential convergence of the likelihood to zero and (b) the high-probability bound on the prior density from Assumption \ref{assumption:prior}.
By the continuous mapping theorem, we similarly have $(2\epsilon_n \sqrt{n})^p / k(\vec{s}^*_n, \vec{y}_{\vec{a}n}) \overset{p}{\to} K^{-1}$.
Combined with the fact that $p\left\{\vec{s}_n \in \mathcal{N}_{\epsilon_n}\left(\vec{s}_n^*\right) | \vec{\theta}, \vec{y}_{\vec{a}n} \right\} / (2\epsilon_n \sqrt{n})^p$ is bounded with high probability by Lemma \ref{lemma:likelihood}, we obtain the following approximation to the posterior density:
\[
p\left\{\vec{\theta} | \vec{y}_{\vec{a}n}, \vec{s}_n \in \mathcal{N}_{\epsilon_n}\left(\vec{s}_n^*\right)\right\} =
K^{-1} \cdot
p(\vec{\theta} | \vec{y}_{\vec{a}n}) \cdot
\frac{p\left\{\vec{s}_n \in \mathcal{N}_{\epsilon_n}\left(\vec{s}_n^*\right) | \vec{\theta}, \vec{y}_{\vec{a}n} \right\}}{(2\epsilon_n \sqrt{n})^p} + o_p(1).
\]
Applying a similar set of steps to the prior and likelihood yields the result in the theorem.


\subsection{Proof of Corollary \ref{corollary:posterior-mean}}\label{proof:cor-pm}

Corollary \ref{corollary:posterior-mean} follows directly from Theorem \ref{thm:bvm} and standard arguments concerning posterior functionals \citep{ghosh_bayesian_2003}.

\subsection{Proof of Corollary \ref{corollary:hodges-lehmann}}\label{sec:proof-cor-hl}

Corollary \ref{corollary:hodges-lehmann} follows directly from the asymptotic expression for $\widetilde{\vec{\theta}}_{n}$, the definition of $\vec{\mu}_n$, and Corollary \ref{corollary:posterior-mean} because
\[\widetilde{\vec{\theta}}_{n} = \vec{\theta}^* + \{\vec{r}'(\vec{\theta}^*)\}^{-1}\{\vec{s}_n - \vec{r}(\vec{\theta}^*)\} + o_p(n^{-1/2}) = \vec{\mu}_n + o_p(n^{-1/2}) = \widehat{\vec{\theta}}_n + o_p(n^{-1/2}).\]

Note that there is some ambiguity in terms of how to define Hodges--Lehmann estimators when there is not a single value $\vec{\theta}$ that solves $\vec{s}_n = \vec{r}_n(\vec{\theta})$;
however, the asymptotic expression $\widetilde{\vec{\theta}}_{n} = \vec{\theta}^* + \{\vec{r}'(\vec{\theta}^*)\}^{-1}\{\vec{s}_n - \vec{r}(\vec{\theta}^*)\} + o_p(n^{-1/2})$ suffices for our purposes.

\subsection{Proof of Theorem \ref{thm:mu_n}}\label{sec:proof-thm-mu_n}

Because $\vec{\mu}_n \coloneqq \vec{\theta}^* + \{\vec{r}'(\vec{\theta^*})\}^{-1} \{\vec{s}_n - \vec{r}(\vec{\theta^*})\}$, Assumption \ref{assumption:clt} immediately implies that $\vec{\mu}_n$ is asymptotically Gaussian.
The expectation is then $\E(\vec{\mu}_n | \vec{y}_{\vec{a}n}) = \vec{\theta}^* + o_p(n^{-1/2})$ by Assumption \ref{assumption:r}.
We further have
\[n \cdot \Var(\vec{\mu}_n | \vec{y}_{\vec{a}n}) = \{\vec{r}'(\vec{\theta^*})\}^{-1}\mat{V}_n(\vec{\theta}^*) \{\vec{r}'(\vec{\theta^*})\}^{-\top} \overset{p}{\to} \{\vec{r}'(\vec{\theta^*})\}^{-1}\mat{V}(\vec{\theta^*})\{\vec{r}'(\vec{\theta^*})\}^{-\top} \eqqcolon \mat{\Sigma}\]
by Assumption \ref{assumption:V}.
The Theorem then follows from an application of Slutsky's Theorem.

\section{Additional Theory Examples}\label{app:additional-theory-examples}

This appendix provides two additional theory examples similar to that of Section \ref{sec:theory-example}.

\subsection{Inverse Probability Weighting}
\label{sec:ipw}

We now consider simple randomization with $a_i \sim \Bernoullidist(\pi_i)$, independently.
We again employ the constant treatment effect model, but we replace $s_{\Delta n}$ with the following inverse-probability-weighted (IPW) statistic: $s_{\text{IPW}n} \coloneqq
\frac{1}{n} \sum_{i=1}^n \widehat{\tau}_{i}$, where $\widehat{\tau}_{i} \coloneqq y_{ai} \left\{a_i/\pi_i - (1 - a_i)/(1 - \pi_i)\right\}$.
Simple computations reveal that $r(\theta) = r_n(\theta) = \theta$, $\theta^* = \E(y_{1i} - y_{0i})$, and $\mu_n = \widetilde{\theta}_n = s_{\text{IPW}n}$.
In turn, we can show that
\[
v_n(\theta) = \frac{1}{n} \sum_{i=1}^n \frac{\{y_{ai} + (1 - \pi_i - a_i)\theta \}^2}{\pi_i (1-\pi_i)}.
\]
In contrast, we have \[
n \cdot \Var(s_{\text{IPW}n} | \vec{y}_{0n}, \vec{y}_{1n}) =
\frac{1}{n} \sum_{i=1}^n \frac{\{(1-\pi_i)y_{1i} + \pi_i y_{0i}\}^2}{\pi_i (1-\pi_i)}.
\]
Substituting $y_{1i} = y_{0i} + \theta^*$ and $y_{ai} = (1-a_i)y_{0i} + a_i y_{1i}$, we see that $v_n(\theta^*) = \Var(s_{\text{IPW}n} | \vec{y}_{0n}, \vec{y}_{1n})$ under correct model specification.
Taking expectations, the form of $v(\theta)$ is given by
\[
v(\theta) = \E \left[\frac{(y_{1i} - \pi_i\theta)^2}{1-\pi_i} + \frac{\{y_{0i} + (1 - \pi_i)\theta\}^2}{\pi_i}\right].
\]
Compared to Section \ref{sec:theory-example}, the relationship between $v(\theta)$ and the frequentist variance is not as straightforward.

\subsection{H\'ajek Estimator}
\label{sec:hajek}

This section considers the setting of Section \ref{sec:ipw} with the H\'ajek estimator:
\[
s_{\text{H}n} \coloneqq
\frac{\sum_{i=1}^n y_{ai} a_i / \pi_i}{\sum_{i=1}^n a_i / \pi_i} -
\frac{\sum_{i=1}^n y_{ai} (1-a_i) / (1-\pi_i)}{\sum_{i=1}^n (1-a_i) / (1-\pi_i)}.
\]
In this case, $r_n(\theta) \neq r(\theta)$ in general due to finite-sample bias, but we do have $r(\theta) = \theta$.
This fact then implies that $\theta^* = \E(y_{1i} - y_{0i})$ and $\mu_n = \widetilde{\theta}_n = s_{\text{H}n}$ as before.
We can show that
\[
v_n(\theta) = 
\frac{1}{n}
\sum_{i=1}^n \frac
    {\left[(1-\pi_i) \left\{ y_{ai} + (1 - a_i)\theta - u_{1n}(\theta) \right\}  + \pi_i \{y_{ai} - a_i \theta - u_{0n}(\theta)\}\right]^2}
    {\pi_i (1 - \pi_i)}
+ o_p(1),
\]
where $u_{1n}(\theta) \coloneqq \Bar{y}_n + n_0 \theta /n$, $u_{0n}(\theta) \coloneqq \Bar{y}_n - n_1 \theta /n$, and $\Bar{y}_n \coloneqq \sum_{i=1}^n y_{ai} / n$.
The finite-population variance of $s_{\text{H}n}$ is
\begin{align}
n \cdot \Var(s_{\text{H}n} | \vec{y}_{0n}, \vec{y}_{1n}) &=\ 
\frac{1}{n} \sum_{i=1}^n \frac{\{(1-\pi_i)(y_{1i} - \Bar{y}_{1n}) + \pi_i (y_{0i} - \Bar{y}_{0n})\}^2}{\pi_i (1-\pi_i)} + o_p(1) \nonumber \\
&\leq\ \frac{1}{n} \sum_{i=1}^n \left\{
\frac{(y_{1i} - \Bar{y}_{1n})^2}{\pi_i} +
\frac{(y_{0i} - \Bar{y}_{0n})^2}{1 - \pi_i}
\right\}+ o_p(1). \label{eq:hajek-var-bound}
\end{align}
Some algebra shows that $y_{1i} - \Bar{y}_{1n} = y_{ai} + (1 - a_i)\theta - u_{1n}(\theta)$ and $y_{0i} - \Bar{y}_{0n} = y_{ai} - a_i \theta - u_{0n}(\theta)$ if $y_{1i} = y_{0i} + \theta$ so that these expressions agree under correct model specification.
If $\pi_i = \pi$ is constant, then $v(\theta^*) = \Var(y_{1i})/(1-\pi) + \Var(y_{0i})/\pi$ because $u_{1n}(\theta^*)$ and $u_{0n}(\theta^*)$ converge in probability to $\E(y_1)$ and $\E(y_0)$, respectively.
Comparing to \eqref{eq:hajek-var-bound}, we see that $v(\theta^*)$ coincides with the frequentist variance bound if $\pi = 0.5$ or $\Var(y_{1i}) = \Var(y_{0i})$---the same conditions as those in Section \ref{sec:theory-example}.

\section{Additional Application Results}\label{app:additional-application-results}

This appendix includes additional graphical results from the application described in Section \ref{sec:case-study}.

\begin{figure}[htbp]
\centering
\begin{subfigure}{\textwidth}
\includegraphics[width=\linewidth]{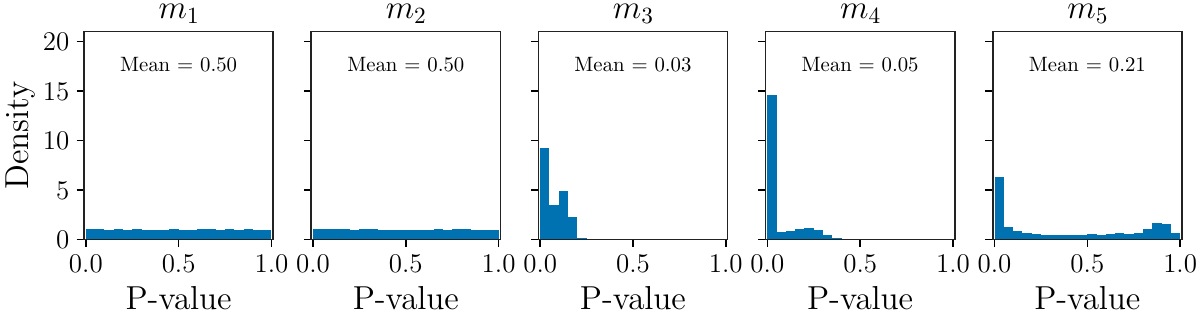}
\caption{Model \eqref{eq:gaussian-reg} Posterior $p$-values}
\label{fig:shift-scale-moments}
\end{subfigure}
\begin{subfigure}{\textwidth}
\includegraphics[width=\linewidth]{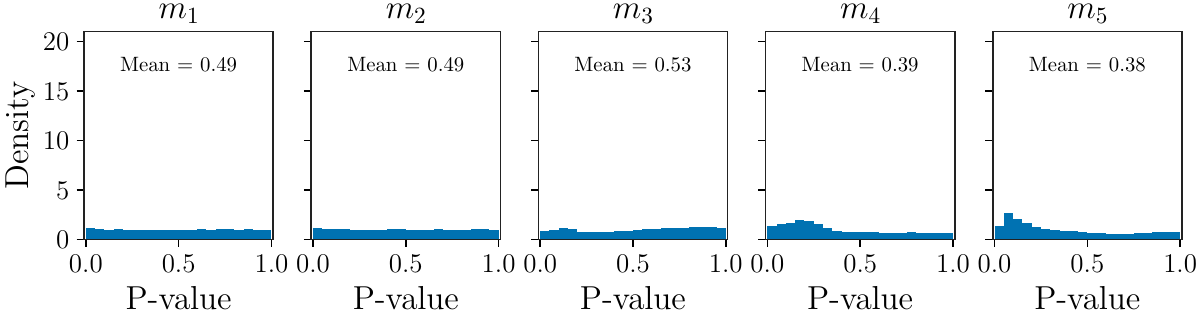}
\caption{Model \eqref{eq:spline-model} Posterior $p$-values}
\label{fig:spline-moments}
\end{subfigure}
\caption{Panel (a) plots posterior predictive checks for the first five centered moments of $y_{1i}$ for model \eqref{eq:gaussian-reg}. Panel (b) plots the same posterior predictive checks for model \eqref{eq:spline-model}.}
\label{fig:moment-p-values}
\end{figure}

Figure \ref{fig:shift-scale-moments} plots posterior $p$-values for the first five centered and scaled moments of the distribution of $y_{1i}$, denoted as $m_1$--$m_5$ in the figure.
Letting $u$ denote the quantile of the observed moment in its randomization distribution, we computed these $p$-values as $2(0.5 - |0.5 - u|)$, effectively giving a two-sided test.
The $p$-values are nearly uniform and have an average value of about 0.5 for $m_1$ and $m_2$, indicating that \eqref{eq:gaussian-reg} adequately models the first two moments of $y_{1i}$.
However, the smaller $p$-values for $m_3$--$m_5$ indicate that model \eqref{eq:gaussian-reg} fails to adequately capture some of the higher-order moments, especially the third and fourth moments.
Figure \ref{fig:spline-moments} plots the posterior predictive $p$-values for this model.
All lie within the range 0.39--0.55, indicating that the first five fitted moments closely match those of the observed data.

\begin{figure}[htbp]
\centering
\begin{subfigure}{0.49\textwidth}
\includegraphics[width=\linewidth]{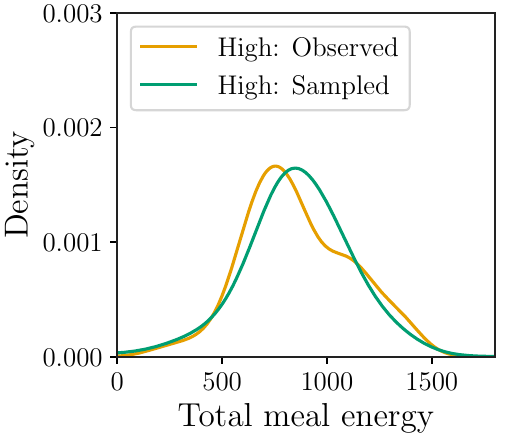}
\caption{Model \eqref{eq:gaussian-reg}}
\label{fig:shift-scale-y1}
\end{subfigure}
\begin{subfigure}{0.49\textwidth}
\includegraphics[width=\linewidth]{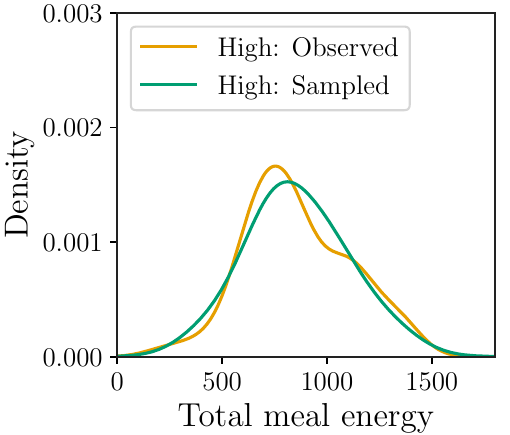}
\caption{Model \eqref{eq:spline-model}}
\label{fig:increasing-y1}
\end{subfigure}
\caption{Panels (a) and (b) plot KDEs for the observed and imputed values of $y_{1i}$ resulting from models \eqref{eq:gaussian-reg} and \eqref{eq:spline-model}, respectively.}
\label{fig:y1}
\end{figure}

Figure \ref{fig:shift-scale-y1} illustrates these higher-order discrepancies in the observed vs. sampled values of $y_{1i}$; in particular, model \eqref{eq:gaussian-reg} does not adequately capture the right skew.
Figure \ref{fig:increasing-y1} shows that this model provides a better fit to the data compared to model \eqref{eq:gaussian-reg}, especially in terms of skew.
The model fit is smoother than the KDE curve, likely due to the smooth polynomial structure and prior regularization.

\section{Extensions}
\label{app:extensions}

This appendix discusses extensions to the basic BRI framework, enabling its application to a wider range of causal inference problems.

\subsection{Covariates}
\label{sec:covariates}

We first extend the setup of Section \ref{sec:framework} to include a vector, $\vec{x}_i \in \R^q$, of pretreatment covariates for each $i \in [n]$.
As in Section \ref{sec:framework}, we arrange these covariates in a matrix, $\mat{X} \in \R^{n \times q}$.
Assumptions \ref{assumption:unconfoundedness} and \ref{assumption:known_assignment_mechanism} can then be weakened as follows.

\begin{assumption}\label{assumption:conditional-unconfoundedness}
    (\emph{Conditional Unconfoundedness}) Conditional on the covariates, the treatment assignments are randomly assigned independent of the potential outcomes: $\vec{a} \indep \mat{Y} | \mat{X}$.
\end{assumption}

\begin{assumption}\label{assumption:conditional_known_assignment_mechanism}
    (\emph{Known Conditional Assignment Mechanism}) The random assignment mechanism, $\PP(\vec{a} | \mat{X})$, is known.
\end{assumption}

In essence, Assumptions \ref{assumption:conditional-unconfoundedness} and \ref{assumption:conditional_known_assignment_mechanism} require that the treatments are randomly assigned within strata determined by $\mat{X}$ and, further, the probability distribution for these treatment assignments is known.
Because these assumptions are weaker than Assumptions \ref{assumption:unconfoundedness} and \ref{assumption:known_assignment_mechanism}, this extension enables the application of BRI to more complex experiments (such as blocked designs) in which treatments may not be marginally independent of the potential outcomes.
The analysis then treats both $\vec{y_a}$ and $\mat{X}$ as fixed as we compute the posterior distribution:
\begin{equation}
\label{eq:bri-covariate-model}
    p(\vec{\theta} | \vec{s}, \vec{y_a}, \mat{X}) \propto
    p(\vec{\theta} | \vec{y_a}, \mat{X})\, p(\vec{s} | \vec{\theta}, \vec{y_a}, \mat{X}).
\end{equation}

Covariates offer two additional benefits compared to the basic framework introduced in Section \ref{sec:framework}.
The first benefit is that they allow us to estimate causal moderation models, such as the linear moderation model: $y_{1i} = y_{0i} + \vec{x}_i^{\top} \vec{\theta}$.
Estimating these models requires richer statistics capable of identifying the additional parameters in $\vec{\theta}$.
For the linear moderation model, for instance, we could use the statistic $\vec{s}_{\text{OLS}} \coloneqq (\mat{X}^{\top}\mat{X})^{-1}\mat{X}^{\top}\widehat{\vec{\tau}}$, where the $i$th entry of $\widehat{\vec{\tau}}$ is $\widehat{\tau}_i$.
A slight modification of the theoretical results in Section \ref{sec:bvm} show that the posterior mean would then be asymptotically equivalent to $\vec{s}_{\text{OLS}}$ with the posterior concentrating around $(\mat{X}^{\top}\mat{X})^{-1}\mat{X}^{\top}\vec{\tau}$, where the $i$th entry of $\vec{\tau}$ is $\tau_i \coloneqq y_{1i} - y_{0i}$.

The second additional benefit of covariates is that they can improve efficiency by removing systematic variation in the outcome.
As an example of the latter, we could modify the setup of Section \ref{sec:ipw}, replacing $\widehat{\tau}_i$ with the following doubly robust (DR) pseudo-outcome similar to those used in \citetApp{bang_doubly_2005, nie_quasi-oracle_2021, kennedy_towards_2023}:
\[
\widehat{\tau}_{\text{DR}i} \coloneqq
\frac{y_{ai} - \vec{x}_i^{\top} \widehat{\vec{\beta}}_{a_i}}{a_i - (1 - \pi_i)} + \vec{x}_i^{\top} (\widehat{\vec{\beta}}_1 - \widehat{\vec{\beta}}_0),
\]
where $\widehat{\vec{\beta}}_0$ and $\widehat{\vec{\beta}}_1$ are estimated regression coefficients for which the estimation error is asymptotically negligible \citepApp[][p. 202]{chen_doubly_2020, wu_sampling_2020}.
By removing variation explained by $\vec{x}_i$, this modification will generally result in lower asymptotic variance compared to the specification in Section \ref{sec:ipw}.

\subsection{Sensitivity Analysis}
\label{sec:sensitivity-analysis}

In the case of observational studies, analysts may desire to explore the robustness of causal findings to Assumptions \ref{assumption:unconfoundedness} and \ref{assumption:conditional-unconfoundedness}.
Within the BRI framework, we can accomplish this goal by assuming a model for $\PP(\vec{a} | \vec{y}_0, \vec{y}_1)$, similar to the Bayesian sensitivity analysis methods described in \citetApp{miller_sensitivity_2000}, \citetApp{steenland_monte_2004}, and \citetApp{greenland_multiple-bias_2005}.
We provide an example below.

\begin{example}
\label{ex:sensitivity-constant}
Assume the constant treatment effect model, $y_{1i} = y_{0i} + \theta$, and $a_i | y_{0i}, y_{1i} \overset{ind}{\sim} \Bernoullidist(\pi_i)$, where $\log\{\pi_i / (1 - \pi_i)\} = \alpha + \beta y_{1i}$.
We could then perform a data analysis using a grid of values for $\alpha, \beta$ to assess sensitivity to varying degrees of confounding.
Alternatively, we could assume a prior distribution for $\alpha, \beta$ and perform an analysis that averages over the uncertainty in their values.
\end{example}

This general setup would also be applicable to a unidirectional model, such as \eqref{eq:gaussian-effects}; though, the model for $\pi_i$ could depend only on the imputed potential outcome ($y_{1i}$ in this case).

\subsection{Estimation of Assignment Mechanism}
\label{sec:assignment-est}

A notable shortcoming of the sensitivity analysis procedures described in Example \ref{ex:sensitivity-constant} is that they do not allow the data to inform the values of the sensitivity parameters, $\alpha$ and $\beta$.
Although $\alpha$ and $\beta$ are not fully identified, the data should allow us to rule out many $(\alpha, \beta)$ pairs that are not consistent with the observed proportions in each group (treatment vs. control).
\citetApp{mccandless_comparison_2017} make a similar point in comparing Bayesian and Monte Carlo sensitivity analyses.

To remedy this issue, we can augment our statistic, $\vec{s}$, to include $\widehat{\pi} \coloneqq n_1/n$.
We could then perform a joint analysis that estimates the full parameter vector, $\vec{\eta} \coloneqq (\alpha, \beta, \vec{\theta}^{\top})^{\top}$.
This approach would enable us to gauge the robustness of causal findings while allowing for patterns of confoundedness that are compatible with the observed data.

In a similar fashion, we could apply this strategy to estimate assignment mechanisms under Assumption \ref{assumption:unconfoundedness} or \ref{assumption:conditional-unconfoundedness}.
For example, suppose we are willing to employ Assumption \ref{assumption:conditional-unconfoundedness} and posit the model $a_i | \vec{x}_i \overset{ind}{\sim} \Bernoullidist(\pi_i)$, $\log\{\pi_i / (1 - \pi_i)\} = \vec{x}_i^{\top} \vec{\beta}$.
Then, as above, we could augment our statistic to include entries that will allow us to estimate $\vec{\beta}$.
In particular, we could compute the maximum likelihood estimator, $\widehat{\vec{\beta}}$, for the assumed logistic regression model and form an enlarged statistic, $\vec{t} \coloneqq (\widehat{\vec{\beta}}^{\top}, \vec{s}^{\top})^{\top}$, to estimate the full parameter vector, $\vec{\eta} = (\vec{\beta}^{\top}, \vec{\theta}^{\top})^{\top}$.

\subsection{Beyond Binary Treatments}
\label{sec:beyond-binary}

Although the main paper considers only binary treatments, the BRI framework can also be applied to richer types of treatment variables, such as continuous treatments or discrete treatments with three or more levels.
In fact, the theory in Section \ref{sec:theory} still applies provided Assumptions \ref{assumption:consistency}--\ref{assumption:clt} are satisfied.
Because deterministic treatment effect models imply values for all counterfactuals, they can be applied in much the same way as described in Section \ref{sec:framework}.

With three or more treatment levels, stochastic treatment effect models consist of models for conditional distributions of the form $\PP(y_{ji} | y_{li})$, where $j$ and $l$ denote distinct treatment levels.
We may always specify a joint distribution for all potential outcomes, resulting in a multidirectional model; however, this approach requires specification of marginal outcome distributions, so the benefit of this approach compared to superpopulation models is unclear.
Alternatively, we may opt to specify a unidirectional model comprising the conditional distributions of only a single potential outcome given each of the others.
For example, with three treatment levels, we could specify $\PP(y_{2i} | y_{1i})$ and $\PP(y_{2i} | y_{0i})$, thereby avoiding the need to specify marginal distributions.
As with binary outcomes, this approach imposes restrictions on the allowable set of statistics; the statistic may involve only those potential outcomes that can be imputed from the model---$\{y_{2i}\}_{i \in [n]}$ in our example---which could prove overly restrictive with many treatment levels.

\newpage
\bibliographystyleApp{apalike}
\bibliographyApp{references}
\end{document}